\documentclass[epj,nopacs,twocolumn,final]{svjour}

\usepackage{graphicx}
\usepackage{amsfonts,amsmath,amssymb}
\usepackage{subcaption}
\usepackage[
    colorlinks,
    linkcolor=blue,
    anchorcolor=black,
    citecolor=blue,
    urlcolor=blue
]{hyperref}
\usepackage[all]{hypcap}
\usepackage{ulem}

\usepackage[
    style=numeric,
    backend=biber,
    sorting=none,
    citestyle=numeric-comp
]{biblatex}
\DeclareUnicodeCharacter{0301}{\'{e}}
\DeclareUnicodeCharacter{0302}{\^{o}}

\newcommand{\alphas}{\ensuremath{\alpha_{\rm s}}}
\newcommand{\alphae}{\ensuremath{\alpha_{\rm em}}}

\newcommand{\dd}{\ensuremath{\mathrm{d}}}

\journalname{Eur. Phys. J. C}

\begin{document}

\title{Momentum and angular correlations in \texorpdfstring{$Z/\gamma$}{Z/gamma}-hadron production in relativistic heavy-ion collisions}

\author{
Zhan Gao \inst{1}
\and Lin Chen \inst{2,3}
\and Peng-Hui Hu \inst{1}
\and Man Xie \inst{4,5}
\and Han-Zhong Zhang \inst{1,4,5}
}

\institute{ 
Key Laboratory of Quark and Lepton Physics (MOE) and Institute of Particle Physics, Central China Normal University, Wuhan 430079, China
\and
School of Science and Engineering, The Chinese University of Hong Kong, Shenzhen 518172, China 
\and
University of Science and Technology of China, Hefei, Anhui, 230026, P.R.China
\and 
Key Laboratory of Atomic and Subatomic Structure and Quantum Control (MOE), Guangdong Basic Research Center of Excellence for Structure and Fundamental Interactions of Matter, Institute of Quantum Matter, South China Normal University, Guangzhou 510006, China
\and 
Guangdong-Hong Kong Joint Laboratory of Quantum Matter, Guangdong Provincial Key Laboratory of Nuclear Science, Southern Nuclear Science Computing Center, South China Normal University, Guangzhou 510006, China
}

\mail{\\
gaozhan2@mails.ccnu.edu.cn\\
raymond.chen88@outlook.com\\
huph@mails.ccnu.edu.cn\\
manxie@m.scnu.edu.cn\\
zhanghz@mail.ccnu.edu.cn}

\abstract{
We carry out a detailed study of medium modifications on momentum and angular correlations between a large transverse momentum hadron and a $Z/\gamma$ trigger in relativistic heavy-ion collisions within a perturbative QCD parton model improved by the Sudakov resummation technique. The total energy loss of a hard parton propagating inside the medium is employed to modify the fragmentation function, while the medium-induced transverse momentum broadening is included in the resummation approach, and both of them are related to the jet transport parameter and obtained by the high-twist formalism.
We obtain good agreements with the existing data on transverse momentum and azimuthal angular correlations for the $Z/\gamma$-hadron pairs in $pp$ and $AA$ collisions, and predict the correlations for the $\gamma$-hadron in central $PbPb$ collisions at 5.02 TeV. The numerical analyses for the $Z/\gamma$-hadron in central $PbPb$ collisions show that the normalized angular distribution is decorrelated due to the medium-induced transverse momentum broadening, however, the angular correlation is enhanced due to the parton energy loss, namely anti-broadening.
The observed modification of the angular correlation is a result of the competition between the broadening and the anti-broadening. This work provides a reliable theoretical tool for a comprehensive and precise study of jet quenching in relativistic heavy-ion collisions.
}

\maketitle


\section{Introduction}\label{sec:intro}
The strongly-coupled state of deconfined quarks and gluons known as the Quark-Gluon Plasma (QGP) is one of the most important discoveries in recent high-energy heavy-ion colliding experiments carried out by the Relativistic Heavy-Ion Collider (RHIC) \cite{Gyulassy:2004zy} and the Large Hadron Collider (LHC) \cite{Muller:2012zq}.
High-energy jets when traversing the strongly-coupled medium, undergo multiple scatterings that induce gluon radiations off the jets.
The jets lose energies and get additional transverse momentum broadening $\langle p_{\perp}^2\rangle$ due to the interactions between the jets and QGP.
This jet quenching \cite{Wang:1992qdg} phenomenon has become a powerful tool for the study of transport properties of the QGP. 
A single parameter called jet transport coefficient ($\hat{q}$) is defined as the average transverse momentum broadening squared per unit length, $\hat{q}=\dd\langle p_{\perp}^2\rangle/\dd L$, which quantifies the strength of momentum exchanges transverse to the direction of jet parton caused by the in-medium scatterings \cite{Baier:1996kr,Baier:1996sk,Baier:1998kq,Liu:2015vna}.
With the assumption of jet energy loss being proportional to the jet transport coefficient,
the values of $\hat{q}$ have been phenomenologically extracted through model-to-data comparisons on hadron suppression \cite{Qin:2007rn,Schenke:2009gb,Chen:2010te,JET:2013cls,Liu:2015vna,Das:2015ana,Feal:2019xfl,Xie:2019oxg,JETSCAPE:2021ehl}. 
Recently, an information field approach is applied to the Bayesian (IF-Bayesian) inference of $\hat{q}$ from combined experimental data on the suppression of single-/di-/$\gamma$-hadrons in heavy-ion collisions at both RHIC and LHC energies \cite{Xie:2022ght,Xie:2022fak}. 
The extracted $\hat{q}/T^3$ is free of long-range correlations, and exhibits a strong temperature ($T$) dependence without an explicit functional parametrization.
Apart from the $\hat{q}$ extraction using hadron suppression due to jet energy loss, the transverse momentum asymmetry of di-/$\gamma$-/$Z$-jets and the angular decorrelation of di-hadrons or hadron-jets are also attempted to constrain the jet transport coefficient due to medium-induced transverse momentum broadening \cite{Chen:2016vem,Chen:2016cof,Chen:2018fqu,Chen:2020kex}.

The perturbative QCD (pQCD) calculation at next-to-leading order (NLO) works well for di-/$\gamma$-hadron transverse momentum spectra except for their angular correlations near the end points in heavy-ion collisions \cite{Zhang:2007ja,Zhang:2009rn}. 
The NLO pQCD expansion fails to converge for the azimuthal angular distribution of the back-to-back di-/$\gamma$-hadron because of the appearance of large Sudakov logarithms. A Sudakov resummation technique is recently developed to make up for the insufficiency \cite{Mueller:2012uf,Mueller:2013wwa,Sun:2014gfa}. 
With the resummation-improved pQCD framework, the angular decorrelation of di-hadrons, hadron-jets and di-jets are obtained and fitted data well due to the vacuum soft gluon radiation (the Sudakov effect) as well as the medium-induced transverse momentum broadening $\langle p_{\perp}^2\rangle$ \cite{Chen:2016vem,Chen:2016cof}. 
The same framework is then applied to study $\gamma/Z$-jet transverse momentum correlations \cite{Chen:2018fqu,Chen:2020kex} in $AA$ collisions, and gluon saturation phenomenon in $pA$ collisions with small-$x$ formalism \cite{Stasto:2018rci,Benic:2022ixp}. 
However, the influences of jet energy loss suppression are absent in the studies of $\langle p_{\perp}^2\rangle$ effects for the back-to-back azimuthal angular correlation of di-hadrons, hadron-jets and di-jets in heavy-ion collisions. 

In jet quenching formalism \cite{Baier:1996kr,Baier:1998kq,Baier:1996sk,Guo:2000nz,Wang:2001ifa}, jet energy loss $\Delta E$ and $\langle p_{\perp}^2\rangle$ effects are two sides of the same coin since $\Delta E\propto\hat{q}=\dd\langle p_{\perp}^2\rangle/\dd L$. 
The former leads to the yield suppression \cite{Gyulassy:1990ye,Wang:1992qdg,Qin:2015srf} and the azimuthal anisotropy \cite{Gyulassy:2000gk} of large transverse momentum hadrons or jets, while the latter means a medium kick transverse to the jets and gives slight enhancements of small transverse momentum hadron spectra \cite{Wang:1998ww} and angular decorrelations in di-/$\gamma$-/$Z$-jet production \cite{Chen:2016vem,Chen:2016cof,Chen:2018fqu,Chen:2020kex}.
In the resummation-improved pQCD framework, the typical value of di-jet transverse momentum imbalance ($\vec q_{\perp AA} = \vec p_{T\rm{jet}1} + \vec p_{T\rm{jet}2}$) in $AA$ collisions is given by a pocket formula, $q_{\perp AA}^{*2} \sim q_{\perp pp}^{*2} + \langle p_{\perp}^2\rangle$, in which $q_{\perp pp}^{*2}$ gives the Sudakov effect from the initial di-jet transverse momentum imbalance and dominates the angular correlations for the large transverse momentum di-jets in high-energy $AA$ collisions at LHC \cite{Chen:2016vem}. 
Since $q_{\perp pp}^{*2}$ is relative to the initial parton energy, parton energy loss effect appearing in the yield for the $q_{\perp AA}^{*2}$ dependence of angular correlation of di-/$\gamma$-/$Z$-jets will result in the energy loss dependence of angular correlation due to the bias of the same final transverse momentum in $pp$ and $AA$ collisions.

In this paper we will present a tomographic study on heavy-ion collisions via transverse momentum and azimuthal angular correlations in {$Z/\gamma$}-hadron production within a pQCD parton model improved by the Sudakov resummation technique. The high-twist (HT) energy loss formalism \cite{Guo:2000nz,Wang:2001ifa} is combined with the pQCD parton model, and the medium-induced transverse momentum broadening $\langle p_{\perp}^2\rangle$ is included into the Sudakov resummation technique \cite{Chen:2016vem,Chen:2018fqu}. We apply the CLVisc 3+1D hydrodynamics model to simulate the dynamical evolution of the QGP medium created in $AA$ collisions \cite{Chen:2017zte,Tachibana:2017syd,Tachibana:2020mtb,JETSCAPE:2020uew}.

The $Z/\gamma$-hadron are perfect probes for studying longitudinal energy loss and transverse momentum broadening since the $Z/\gamma$ bosons do not interact strongly with the hot and dense medium. Firstly, they do not have geometric bias compared to dihadrons and dijets, therefore, the associated jets can probe more inner regions of the QGP fireball and bring out more robust information about it \cite{Zhang:2007ja,Zhang:2009rn}. Secondly, $Z/\gamma$ bosons could perfectly preserve the initial momentum information of the away-side jets before getting quenched, thus, we can cleanly investigate the relation between the suppression or azimuthal angular correlation and the energy of the initial jet. Finally and importantly, as mentioned above, the longitudinal energy loss effect appears in the angular correlation of di-particles. To reveal the medium modification essentially, we should adopt the same kinematics cuts for final state particles in $pp$ and $AA$ collisions, hence the momentum information for $Z/\gamma$ is the same in $pp$ and $AA$ collisions. Meanwhile, the initial momentum of the associated hadron is larger in $AA$ collisions due to energy loss \cite{Baier:2001yt,Salgado:2003gb,Arleo:2017ntr,Spousta:2015fca}, which causes an anti-broadening effect. In our studies, we will perform a comprehensive and precise study on the anti-broadening effect resulting from the jet energy loss and the broadening effect \cite{Wang:2003mm} caused by medium-induced transverse momentum broadening by calculating the momentum ($p_{\rm T}^h / z_{\rm T} / \xi_{\rm T}$) and angular correlations ($\Delta \phi$) of $Z / \gamma$-hadron production in relativistic heavy-ion collisions.

The paper is organized as follows. 
In \autoref{sec:theory}, we describe the theoretical framework used in our calculations. 
In \autoref{sec:result}, we present and discuss the medium modifications on momentum and angular correlations between a large transverse momentum hadron and the $Z/\gamma$ trigger in $AA$ collisions at RHIC and LHC. 
We end the paper with a summary in \autoref{sec:summary}. 


\section{Theoretical framework}\label{sec:theory}

\subsection{a Sudakov resummation improved pQCD parton model in \texorpdfstring{$pp$}{pp} collisions}
In high-energy $pp$ collisions, the invariant cross section for single inclusive hadron production at large transverse momentum $p_{\rm T}$ within the parton model can be factorized into a convolution of collinear parton distribution functions (PDFs), short-distance partonic cross-sections and the collinear fragmentation functions (FFs).
The $Z/\gamma$-hadron production cross-section within the Sudakov resummation formalism in $pp$ collisions can be expressed as,
\begin{eqnarray}\label{eq:master_pp}
\frac{\dd\sigma_{pp\rightarrow Vh}}{\dd\mathcal{P.S.}}
&=& \sum_{a,b,d} \int \frac{\dd z_d}{z^2_{d}}  D_{h / d}\left(z_{d}, \mu^{2}\right) \frac{|\bar{\mathcal{M}}_{ab\rightarrow Vd}|^2}{16\pi ^2 \hat{s}^2}\nonumber\\
&\times& x_{a} f_{a / p}\left(x_{a}, \mu^{2}\right)  x_{b}f_{b / p}\left(x_{b}, \mu^{2}\right)  \nonumber\\
&\times& \int \frac{\dd^{2} \vec{b}_{\perp}}{(2 \pi)^{2}} e^{-i \vec{q}_{\perp} \cdot \vec{b}_{\perp}} e^{-S(Q,b_{\perp})}.
\end{eqnarray}
Here $\dd \mathcal{P.S.} = \dd y_V~\dd y_h~\dd^{2} \Vec{p}_{\rm T}^{V}~\dd^{2} \Vec{p}_{\rm T}^{h}$ represents the final state phase space, $y_V$ and $y_h$ are the rapidities of the vector boson and hadron, respectively. $p_{\rm T}^V$ and $p_{\rm T}^{h}$ are the transverse momentum of the vector boson and hadron. Initial parton momentum fractions are defined as $x_{a,b} = (m_{\rm T}^V e^{\pm y_V} + m_{\rm T}^d  e^{\pm y_h} )/\sqrt{s}$, with $m_{\rm T} = \sqrt{m^2 + p_{\rm T}^2}$ the transverse mass of the particle, with $m$ the particle mass and $p_{\rm T}$ its transverse momentum. While $\sqrt{s}$ denotes the usual collision energy in the proton-proton Center-of-Mass frame. $f_{a,b}(x,\mu^2)$ are the PDFs of the incoming parton species $a$ and $b$, which we take from Cteq (CT18) parametrizations \cite{Hou:2019efy}. 
$z_d = p_{\rm T}^h/p_{\rm T}^d$ is the momentum fraction of the final observed hadron $h$ carried from the parent parton $d$, and we use the Kretzer (KRE) parametrizations \cite{Kretzer:2000yf} for the vacuum fragmentation function $D_{h / d}\left(z_{d}, \mu^{2}\right)$. 
$\hat{s}=x_a x_b s=Q^2$, $\hat{t}=m_V^2 - x_a m_{\rm T}^V \sqrt{s} e^{-y_V}$ and $\hat{u}=-x_a m_{\rm T}^d \sqrt{s} e^{-y_h}$  are then the usual partonic Mandelstam variables.
The auxiliary $b_{\perp}$-space integral guarantees transverse momentum conservation of the radiated gluons, with $\vec{q}_{\perp} \equiv \vec{p}_{\rm T}^V+\vec{p}_{\rm T}^h/z_{d}=\vec{p}_{\rm T}^V+\vec{p}_{\rm T}^d$ defined as the transverse momentum imbalance of the two out-going final particles. $\Delta\phi = |\phi_{V}-\phi_{h}|$ is the azimuthal
angle difference between the vector boson and the hadron.
The amplitudes squared, averaged over the color and spin in the initial state and summed over them in the final state, are given by
\begin{align}
|\bar{\mathcal{M}}_{q \bar{q} \rightarrow V g}|^2 &=\frac{N_c^2-1}{N_c^2} 16 \pi^2 \alphas \alphae e_q^2 \frac{\hat{t}^2+\hat{u}^2+2 \hat{s} m_V^2}{\hat{t} \hat{u}}, \\
|\bar{\mathcal{M}}_{q g \rightarrow V q}|^2 &=-\frac{1}{N_c} 16 \pi^2 \alphas \alphae e_q^2 \frac{\hat{s}^2+\hat{t}^2+2 \hat{u} m_V^2}{\hat{s} \hat{t}}, 
\end{align}
where $e_q$ is the electric charge of the quarks in the case of photon production. For $Z$ boson production, $e_q$ is replaced by
\begin{equation}
e_q^2 \rightarrow \frac{\left(1-2\left|e_q\right| \sin ^2 \theta_W\right)^2+4 e_q^2 \sin ^4 \theta_W}{8 \sin ^2 \theta_W \cos ^2 \theta_W},
\end{equation}
with $\theta_W$ the weak mixing angle. The values of the $Z$ boson masses $m_{Z}$ and $\theta_W$ are taken from PDG \cite{Workman:2022ynf}. 
Replacing $D_{h / d}\left(z_{d}, \mu^{2}\right)$ to $\delta(1-z_d)$ and performing the integration of $\dd y_h$ and $\dd^{2} \Vec{p}_{\rm T}^h$, we can obtain the differential cross-sections of $Z/\gamma$ bosons production in $pp$ collisions.

The Sudakov form factor is expressed as follows,
\begin{equation}\label{eq:sud_vac}
S_{\text{vac}}(Q, b_\perp) = S^i_{\text{pert}}(Q, b_\perp)  + S^f_{\text{pert}}(Q, b_\perp)  +  S_{\text{NP}}(Q, b_\perp).
\end{equation}
$ S^i_{\text{pert}}(Q, b_\perp) $ is the perturbative Sudakov factor of initial parton, $S^f_{\text{pert}}(Q, b_\perp) $ is perturbative Sudakov factor of final state hadrons, and $ S_{\text{NP}}(Q, b_\perp) $ is the non-perturbative Sudakov factor in $pp$ collisions. 
The perturbative Sudakov factors are defined as \cite{Collins:1984kg,deFlorian:2000pr,Catani:2000vq,deFlorian:2001zd} 
\begin{equation}\label{eq:sud_p}
S^i_{\text{pert}}(Q, b_\perp)
=
\int ^{\mu ^2 _{\rm res}}_{\mu ^2 _{\rm fac}} \frac{\dd \mu ^2 }{ \mu ^2 }
\left[ \left( A_i^{(1)} + A_i^{(2)}    \right) \ln\frac{Q^2}{\mu ^2} +B_i \right],
\end{equation}
\begin{equation}
S^f_{\text{pert}}(Q, b_\perp)
=
 S^i_{\text{pert}}(Q, b_\perp),
\end{equation}
and
\begin{center}
\begin{tabular}{|c|c|c|}\hline
	&	Quark	&	Gluon\\\hline
$A_i^{(1)}$	&	$C_F\left[\frac{\alphas(\mu^2)}{2\pi}\right]$	&	$C_A\left[\frac{\alphas(\mu^2)}{2\pi}\right]$\\
$A_i^{(2)}$	&	$K~C_F \left[\frac{\alphas(\mu^2)}{2\pi}\right]^2$	&	$K~C_A \left[\frac{\alphas(\mu^2)}{2\pi}\right]^2$\\
$B_{i}$	&	$-\frac{3}{2}C_F\left[\frac{\alphas(\mu^2)}{2\pi}\right]$	&	$-2\beta_0C_A\left[\frac{\alphas(\mu^2)}{2\pi}\right]$\\
\hline
\end{tabular}
\end{center}
with  $N_c=3$, $C_A=N_c$, and $C_F=\frac{N_c^2-1}{2N_c}$. $K=\left(\frac{67}{18}-\frac{\pi^2}{6}\right)C_A-\frac{10}{9}N_fT_R$ is the coefficient for the $\alphas^2$ term. To ensure the $q_{\perp} > \Lambda_{\mathrm{QCD}}$ and  to separate the Sudakov form factor into perturbative and non-perturbative parts in $pp$ collisions, the $b_{\star} =b_{\perp}/\sqrt{1+b_{\perp}^2/b_{\max}^2}$ prescription is introduced \cite{Davies:1984sp,Ladinsky:1993zn,Landry:2002ix,Prokudin:2015ysa,Sun:2014dqm}, with $b_{\max}=1.5~\mathrm{GeV^{-1}}$. 
The result is the definition of the factorization scale $\mu_{\rm fac} = b_0/b_{\star}$ with $b_0 =2e^{-\gamma_E}$ and $\gamma_E$ the Euler constant. 
$ S_{\text{NP}}(Q, b_\perp) $ is non-perturbative Sudakov factor in $pp$ collisions. 
It is defined as 
\begin{align}\label{eq:sud_np}
S^q_{\text{NP}}(Q, b_\perp)&=\frac{g_1}{2}b_1^2 +\frac{g_2}{2} \ln \frac{Q}{Q_0} \ln \frac{b_\perp}{b_\star},    \\
S^g_{\text{NP}}(Q, b_\perp)&=\frac{C_A}{C_F}S^q_{\text{NP}}(Q, b_\perp),
\end{align}
where $g_1 =0.212$, $g_2 = 0.84$, and $Q_0^2=2.4$ GeV$^2$
 \cite{Sun:2014dqm}. 

There are three distinct scales in this formalism. The factorization scale $\mu_{\rm fac}$ appearing in the PDF and FF is fixed as $\mu_{\rm fac} = b_0/b_{\star}$. 
The renormalization scale $\mu_{\rm ren}$ in $\left|\Bar{\mathcal{M}}\right|$ is taken to be $\mu_{\rm ren} = m_{\rm T}^V$ \cite{Sun:2018icb}. 
To minimize the contributions from these possible large logs, we will set the resummation scale to $\mu_{\rm res} = p_{\rm T}^d$ \cite{Sun:2016kkh}. 

\subsection{parton energy loss and transverse momentum broadening in \texorpdfstring{$AA$}{AA} collisions}
The cross section of $Z/\gamma$-hadron production in $AA$ collisions can be expressed as,
\begin{equation}\label{eq:master_aa}
		\begin{aligned}
		\frac{\dd \sigma_{AA\rightarrow Vh}}{\dd \mathcal{P.S.}}=& \int \dd^2\Vec{b} \dd^2 \Vec{r}~ t_A(\vec{r}) t_B(\vec{r}+\vec{b})\int \frac{\dd \phi_d}{2 \pi}\\
		\times&\sum_{a b d} \int  \frac{\dd z_{d}}{z^2_{d}}  \tilde{D}_{h / d}\left(z_d, \mu^2, \Delta E_d\right) \frac{|\bar{\mathcal{M}}_{ab\rightarrow Vd}|^2}{16\pi ^2 \hat{s}^2}\\
		\times& x_{a} \tilde{f}_{a / A}\left(x_{a}, \mu^{2},\vec{r} \right )  x_{b}\tilde{f}_{b / A}\left(x_{b}, \mu^{2},\vec{r}+\vec{b}\right)  \\
		\times& \int \frac{\dd^{2} \vec{b}_{\perp}}{(2 \pi)^{2}} e^{-i \vec{q}_{\perp} \cdot \vec{b}_{\perp}} e^{-S(Q,b_{\perp})}.
	\end{aligned}
\end{equation}
Where $\phi_d$ is the azimuthal angle between the parton’s propagating direction and the impact-parameter $\vec{b}$. $t_A$ and $t_B$ are the nuclear thickness functions of the projectile and target nucleus, shifted by the impact parameter $\vec{b}$. 
\footnote{$\vec{b}$ is dependent on the collision centrality, whereas $\Vec{b}_\perp$ is the Fourier conjugate to $\Vec{q}_{\rm T}$ in the Sudakov integral.}
They are calculated by integrating the Woods-Saxon nuclear density distribution \cite{Jacobs:2000wy} along the beam direction and is normalized to the mass number A of each nucleus, e.g., $\int \dd^2
\Vec{r}~t_A(\vec{r}) = {\rm A}$. 
$\tilde{f}_{a / A}\left(x_{a}, \mu^{2},\vec{r} \right )$ is the nuclear modified PDF \cite{Wang:1996yf,Li:2001xa}: 
\begin{equation}\label{eq:npdf}
\begin{aligned}
\tilde{f}_{a / A}\left(x_a, \mu^2, \vec{b}\right) & =R_{a / A}\left(x_a, \mu^2, \vec{b}\right)\left[\frac{\rm Z}{\rm A} f_{a / p}\left(x_a, \mu^2\right)\right. \\
& \left.+\left(1-\frac{\rm Z}{\rm A}\right) f_{a / n}\left(x_a, \mu^2\right)\right],
\end{aligned}
\end{equation}
where Z and A are the charge and mass number of the nucleus, respectively. The cold nuclear modification factor $R_{a / A}\left(x_a, \mu^2, \vec{b}\right)$ of the PDFs are given by the EPPS21 \cite{Eskola:2021nhw} parametrizations. The medium-modified fragmentation function $\tilde{D}_{h / d}\left(z_d, \mu^2, \Delta E_d\right)$ can be calculated as, \cite{Zhang:2007ja,Wang:2004yv,Zhang:2009rn}
\begin{equation}\label{eq:nff}
\begin{aligned}
&\tilde{D}_{h / d}\left(z_d, \mu^2, \Delta E_d\right)=\left(1-e^{-N_g^d}\right)\left[\frac{z_d^{\prime}}{z_d} D_{h / d}\left(z_d^{\prime}, \mu^2\right)\right. \\
&\left.\quad+N_g^d \frac{z_g^{\prime}}{z_d} D_{h / g}\left(z_g{ }^{\prime}, \mu^2\right)\right]+e^{-N_g^d} D_{h / d}\left(z_d, \mu^2\right),
\end{aligned}
\end{equation}
where ${z_d}'=p_{\rm T}^{h}/(p_{\rm T}^{d}-\Delta{E_d}$) is the transverse momentum fraction of a hadron from the fragmentation of parton $d$ with initial transverse momentum $p_{\rm T}^{d}$ after losing a total amount of energy $\Delta{E_d}$ in the medium,
$z_d$ is the momentum fraction of a hadron from a parton fragmentation in vacuum, and ${z_g}'=N_g^d p_{\rm T}^{h}/\Delta{E_d}$ is the momentum fraction of a hadron from the fragmentation of a radiated gluon that carries an average energy $\Delta{E_d}/N_g^d$. The number of radiated gluons is assumed to follow the Poisson distribution with the number $N_g^d$. In this case, the first weighting factor $1-e^{-N_g^d}$ in the above equation is the probability for a parton to radiate at least one gluon induced by multiple scatterings. The second factor $e^{-N_g^d}$ is the probability of no induced gluon radiating.

Within the high-twist (HT) formalism \cite{Wang:2001ifa,Wang:2002ri,Wang:2001cs,Deng:2009ncl}, the radiative energy loss for a parton $d$ with initial energy $E$ can be calculated as,
\begin{equation}
\label{eq:enloss}
\frac{\Delta E_d}{E} = \int \dd \tau \int \dd  l_{\rm T}^2 \int \dd z \left( z \frac{\dd N^{\rm med}_d}{\dd z \dd l_{\rm T}^2  \dd \tau } \right),
\end{equation}
and the number of radiated gluons $N_g^d$ from the propagating parton $d$ is
\begin{equation}
N_g^d = \int \dd \tau\int \dd  l_{\rm T}^2 \int \dd z  \frac{\dd N^{\rm med}_d}{\dd z \dd l_{\rm T}^2  \dd \tau },
\end{equation}
where the medium-induced gluon spectrum is given by \cite{Deng:2009ncl}
\begin{equation}\label{eq:med_gluon_spectrum}
\begin{aligned}
&\frac{\dd N^{\rm med}_d}{\dd z \dd l_{\rm T}^2  \dd \tau } 
= \frac{2 C_A \alphas}{ \pi l_{\rm T}^4} \left[ \frac{1+(1-z)^2}{z}\right] \\
&\times \hat{q}_d\left(\tau, \vec{r}+\left(\tau-\tau_0\right) \vec{n}\right) \sin ^2\left[\frac{l_{\rm T}^2\left(\tau-\tau_0\right)}{4 z(1-z) E}\right].
\end{aligned}
\end{equation}
The integration over the quark propagation path starts from the initial transverse position $\vec r$ at an initial time $\tau_0=0.6$ fm/$c$ along the direction $\vec n$. 
$\alphas$ is the strong coupling constant, $l_{\rm T}$ is the transverse momentum of the radiated gluon, and $z$ is its longitudinal (along the jet direction) momentum fraction. 
The parton energy loss for a propagating gluon is assumed to be the same as the quark, except that the jet transport coefficient of a gluon $\hat{q}_A$ differs from that of a quark $\hat{q}_F$ by the color factor $C_A/C_F$. Therefore, the radiative energy loss of a gluon is $9/4$ times that of a quark \cite{Deng:2009ncl}.

Using information field-based global Bayesian inference, jet transport coefficient $\hat{q}$ is extracted from experimental data of the hadron suppression \cite{Xie:2022fak,Xie:2022ght}. In this work, we only consider parton energy loss in the QGP phase with a pseudo-critical temperature $T_c = 0.165$ GeV. Effectively, we assume the jet transport coefficient $\hat{q}$ in the local co-moving frame as $\hat{q}=\hat{q}(T) p^\mu \cdot u_\mu / p^0$ that depends on the local temperature and vanishes below $T_c$, where $p^\mu=\left(p_0, \vec{p}\right)$ is the four-momentum of the parton and $u^{\mu}$ is the four flow velocity of the fluid. The dynamical evolution of the QGP medium created in $AA$ collisions is provided by the CLVisc $3+1$-D hydrodynamics model simulations \cite{Pang:2012he,Pang:2014ipa,Pang:2018zzo}.

There are two separate contributions to the azimuthal correlation in heavy-ion collisions: one is the vacuum parton shower, which is referred to as the Sudakov effect; the other is the $p_{\rm T}$-broadening effect (or medium effect) due to multiple scatterings and medium-induced radiation when high-energy partons propagate through the medium \cite{Mueller:2012uf,Mueller:2013wwa,Mueller:2016gko,Mueller:2016xoc}. These two effects contribute differently to the transverse momentum broadening in well-separated regions of their phase space integral.
In $AA$ collisions, the transverse momentum broadening is calculated by combining the vacuum Sudakov factor $S(Q, b_{\perp})$ with the medium broadening factor \cite{Chen:2016vem}
\begin{equation}\label{eq:sud_med}
	S_{\mathrm{med}}(Q,b_{\perp})= S_{\mathrm{vac}}(Q,b_{\perp})+ \frac{b_{\perp}^2}{4}\left\langle p_{\perp}^2 \right\rangle_{\rm tot} , 
\end{equation}
where $\left\langle p_{\perp}^2 \right\rangle$ indicates the amount of the medium-induced broadening experienced by partons, and $\left\langle ...\right\rangle $ means averaging over all different paths.
The medium-induced broadening for quark and gluon jets are related as $\left\langle p_{\perp}^2 \right\rangle _{g} = \frac{C_A}{C_F}\left\langle p_{\perp}^2 \right\rangle _{q}$.

An energetic parton propagating through the QGP medium can encounter $2 \rightarrow 2$ elastic scatterings and exchange transverse momentum with the medium. These ``elastic collisions" or ``leading order" contribute to transverse momentum broadening of the hard parton \cite{Majumder:2007hx,Qin:2012fua,Clayton:2021uuv}. Such elastic-scattering contribution to the averaged transverse momentum broadening squared is given by
\begin{equation}
    \left\langle p_{\perp}^2 \right\rangle_{\rm 
 el} =\left\langle \int \dd \tau \hat{q}_d\left(\tau, \vec{r}+\left(\tau-\tau_0\right) \vec{n}\right) \right\rangle.
    \label{p_perp_el}
\end{equation}
 ``Radiative corrections" due to the medium inducing soft gluon radiation also lead to the $p_{\rm T}$-broadening by the recoil effects in the emission of soft gluons \cite{Liou:2013qya,Blaizot:2014bha,Zakharov:2020sfx}. We can obtain $\left\langle p_{\perp}^2 \right\rangle_{\rm rad}$ from \autoref{eq:med_gluon_spectrum} by calculating the mean-square momentum $l_{\rm T}$ of radiated gluons:
 \begin{equation}
    \left\langle p_{\perp}^2 \right\rangle_{\rm rad} =\left\langle \int d\tau \int \dd  l_{\rm T}^2 \int \dd z  \left(l_{\rm T}^2 \frac{\dd N^{\rm med}_d}{\dd z \dd l_{\rm T}^2  \dd \tau } \right) \right\rangle.
    \label{p_perp_rad}
\end{equation}
As an approximation, we assume that the total contributions to $\left\langle p_{\perp}^2 \right\rangle$ are given by the sum of elastic-collision and radiation processes, 
\begin{equation}
\left\langle p_{\perp}^2 \right\rangle_{\rm tot}
= \left\langle p_{\perp}^2 \right\rangle_{\rm el}+ \left\langle p_{\perp}^2 \right\rangle_{\rm rad}. 
\label{eq:p_med_tot}
\end{equation}

\begin{figure}[!ht]
\centering
\includegraphics[width=1\linewidth]{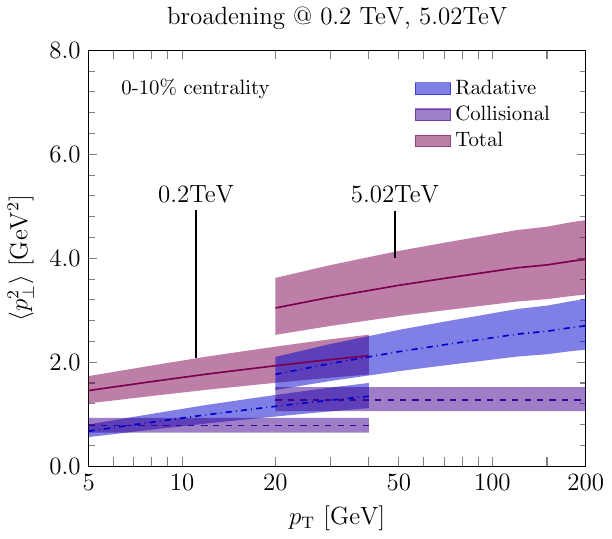}
\caption{(Color online) Averaged transverse momentum broadening for a quark with the median $\hat{q}$ (curves) with uncertainties (bands) as a function of the initial jet transverse momentum $p_{\rm T}$ in 0-10\% $Au+Au$ collisions at $\sqrt{s_{\rm NN}} = 200$ GeV and in 0-10\% $Pb+Pb$ collisions at $\sqrt{s_{\rm NN}} = 5.02$ TeV. The solid, dot-dashing and dashing curves are for the total, radiation and elastic collision contributions, respectively. }
\label{fig:br200}
\end{figure}

We plot the $\left\langle p_{\perp}^2 \right\rangle$ as a function of the partron transverse momentum in \autoref{fig:br200} for 0-10\% $Au+Au$ collisions at $\sqrt{s_{\rm NN}} = 200 \mathrm{GeV}$ and 0-10\% $Pb+Pb$ collisions at $\sqrt{s_{\rm NN}} = 5.02 \mathrm{TeV}$, respectively. Numerical results show that the radiation contribution is larger than the collision contribution, and the total contribution increases slightly with the initial jet transverse momentum at both RHIC and LHC. 


\section{Numerical results and discussions}\label{sec:result}

With the theoretical framework described in \autoref{sec:theory}, we will present the transverse momentum ($p_{\rm T}$ or $z_{\rm T}$) and azimuthal angular ($\Delta \phi$) correlation distributions for $Z/\gamma$-hadron production at RHIC/LHC in details. The production suppression and azimuthal angular decorrelation of $Z/\gamma$-hadron caused by energy loss and transverse momentum broadening will be discussed numerically. At last we calculate the root-mean-square (RMS) width of $\Delta \phi$ distribution to analyze the medium broadening effects.

The per-trigger yield of $Z/\gamma$-hadron production or the $Z/\gamma$-triggered fragmentation function in $pp$ collisions is calculated as 
\begin{equation}
\begin{split}
D_{pp}(p_{\rm T}^h, \Delta\phi)&=\frac{1}{N_V^{pp}} \frac{\mathrm{d}^2 N_{Vh}^{pp}}{\dd p_{\rm T}^h \dd \Delta \phi}\\
&=\frac{ \int \mathrm{d} p_{\rm T}^V (\dd^3 \sigma^{pp}_{Vh} / \dd p_{\rm T}^V \dd p_{\rm T}^h \dd \Delta\phi)}{\int \mathrm{d} p_{\rm T}^V (\dd \sigma^{pp}_{V} / \dd p_{\rm T}^V)},
\end{split}
\end{equation}
where the numerator is $Z/\gamma$-hadron cross section and the denominator is the total cross section of $Z/\gamma$ production. Similarly in $AA$ collisions, the per-trigger yield of $Z/\gamma$-hadron production is given by
\begin{equation}
\begin{split}
D_{AA}(p_{\rm T}^h, \Delta\phi)&=\frac{1}{N_V^{AA}} \frac{\mathrm{d}^2 N_{Vh}^{AA}}{\dd p_{\rm T}^h \dd \Delta \phi}\\
&=\frac{ \int \mathrm{d} p_{\rm T}^V (\dd^3 \sigma^{AA}_{Vh} / \dd p_{\rm T}^V \dd p_{\rm T}^h \dd \Delta\phi)}{\int \mathrm{d} p_{\rm T}^V (\dd \sigma^{AA}_{V} / \dd p_{\rm T}^V)}.
\end{split}
\end{equation}
The nuclear modification factor $I_{\rm AA}$ is defined as the ratio of $Z/\gamma$-hadron yields between $AA$ and $pp$ collisions,
\begin{equation}\label{eq:iaa}
I_{AA}(p_{\rm T}^h, \Delta\phi)
=\frac{D_{pp}(p_{\rm T}^h, \Delta\phi)}{D_{AA}(p_{\rm T}^h, \Delta\phi)}.
\end{equation}
Defining $z_{\rm T} =p_{\rm T}^h/p_{\rm T}^V$, we get the $Z/\gamma$-triggered fragmentation function as \cite{Wang:2003mm}
\begin{equation}
D_{pp}(z_{\rm T}, \Delta\phi)
=\frac{ \int \dd p_{\rm T}^V (p_{\rm T}^V \dd^3 \sigma^{pp}_{Vh} / \dd p_{\rm T}^V \dd p_{\rm T}^h \dd \Delta\phi)}{\int \dd p_{\rm T}^V ( \dd \sigma^{pp}_{V} / \dd p_{\rm T}^V)}.
\end{equation}
In $AA$ collisions, the $Z/\gamma$-triggered fragmentation function is given as
\begin{equation}
D_{AA}(z_{\rm T}, \Delta\phi)
=\frac{ \int \dd p_{\rm T}^V (p_{\rm T}^V \dd^3 \sigma^{AA}_{Vh} / \dd p_{\rm T}^V \dd p_{\rm T}^h \dd \Delta\phi)}{\int \dd p_{\rm T}^V ( \dd \sigma^{AA}_{V} / \dd p_{\rm T}^V)}.
\end{equation}
Then the nuclear modification factor is also written as
\begin{equation}
I_{AA}(z_{\rm T}, \Delta\phi) = \frac{D_{AA}(z_{\rm T}, \Delta\phi)}{D_{pp}(z_{\rm T}, \Delta\phi)}.
\end{equation}
One can deifne $\xi_{\rm T}=\ln (-\left|\vec{p}_{\rm T}^{V}\right|^2 /\vec{p}_{\rm T}^{h} \cdot \vec{p}_{\rm T}^{V})$ \cite{CMS:2021otx}, then get the $Z/\gamma$-triggered fragmentation function in $pp$ collisions as
\begin{equation}
D_{pp}(\xi_{\rm T}, \Delta\phi)
=-\frac{ \int \dd p_{\rm T}^V (p_{\rm T}^h \dd^3 \sigma^{pp}_{Vh} / \dd p_{\rm T}^V \dd p_{\rm T}^h \dd \Delta\phi)}{\int \dd p_{\rm T}^V ( \dd \sigma^{pp}_{V} / \dd p_{\rm T}^V)},
\end{equation}
and in $AA$ collisions as
\begin{equation}
D_{AA}(\xi_{\rm T}, \Delta\phi)
=-\frac{ \int \dd p_{\rm T}^V (p_{\rm T}^h \dd^3 \sigma^{AA}_{Vh} / \dd p_{\rm T}^V \dd p_{\rm T}^h \dd \Delta\phi)}{\int \dd p_{\rm T}^V ( \dd \sigma^{AA}_{V} / \dd p_{\rm T}^V)}.
\end{equation}
The corresponding nuclear modification factor is then written as
\begin{equation}
I_{AA}(\xi_{\rm T}, \Delta\phi) = \frac{D_{AA}(\xi_{\rm T}, \Delta\phi)}{D_{pp}(\xi_{\rm T}, \Delta\phi)}.
\end{equation}

\begin{figure}[!ht]
\centering
\includegraphics[width=1\linewidth]{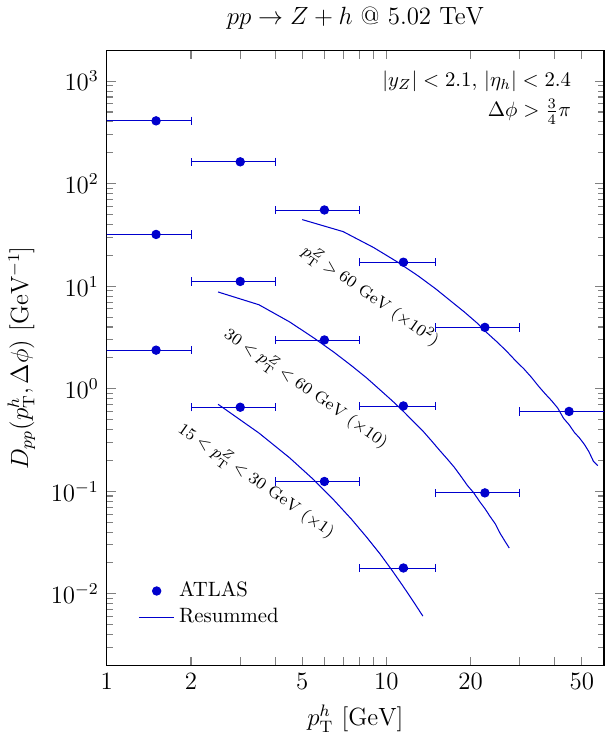}
\caption{(Color online) The $p_{\rm T}^{h}$ distributions of per $Z$-trigger hadrons with three $p_{\rm T}^Z$ ranges are calculated in $pp$ collisions at $\sqrt{s_{\rm NN}}=5.02$ TeV within the Sudakov resummation improved pQCD parton model, and compared with the ATLAS experimental data \cite{ATLAS:2020wmg}.}
\label{fig:ATLAS_pp}
\end{figure}

\subsection{transverse momentum correlations of  \texorpdfstring{$Z/\gamma$}{Z/gamma}-hadrons}

\begin{figure*}[!ht]
\centering
\includegraphics[width=0.8\linewidth]{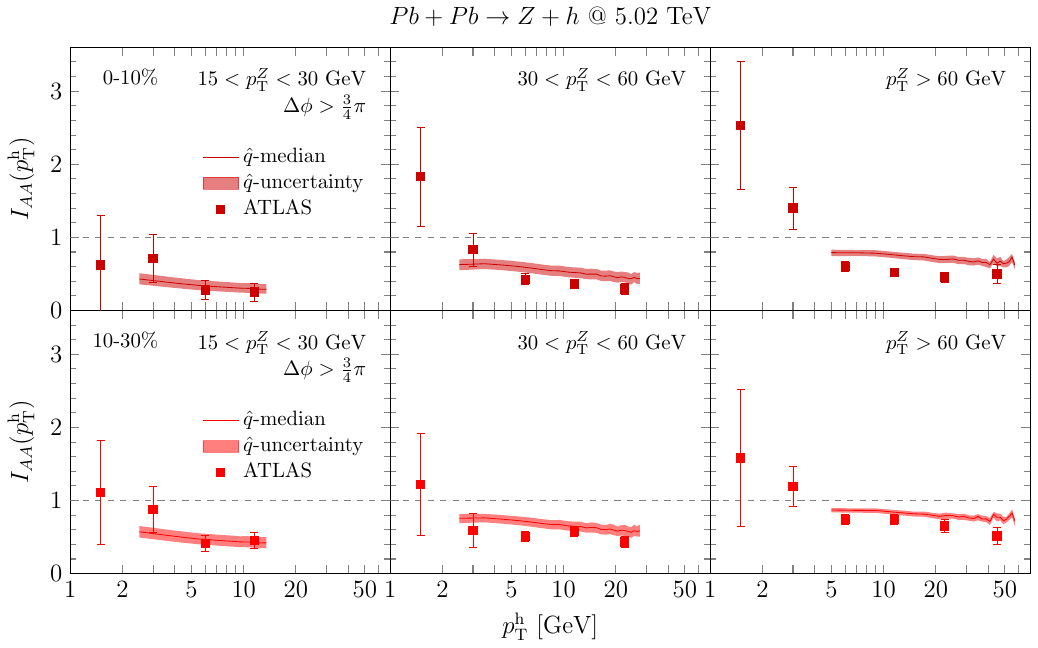}
\caption{(Color online) The nuclear modification factors $I_{AA}$ of $Z$-hadron as a function of $p_{\rm T}^{h}$ with three trigger $p_{\rm T}^Z$ ranges in 0-10\% (upper panels) and 10-30\% (lower panels) $PbPb$ collisions at $\sqrt{s_{\rm NN}}=5.02$ TeV, compared with the ATLAS experimental data \cite{ATLAS:2020wmg}, respectively.}
\label{fig:ATLAS_iaa}
\end{figure*}

Taking the energy loss and transverse momentum broadening into account simultaneously, we first calculate the transverse momentum distribution for $Z/\gamma$-hadron production and show their medium modifications. The numerical results with the Sudakov resummation improved pQCD parton model are compared with all available experimental data at the LHC.

Shown in \autoref{fig:ATLAS_pp} is the transverse momentum spectra of the charged hadrons triggered by $Z$ bosons in $pp$ collisions at $\sqrt{s_{\rm NN}}=5.02$ TeV. The numerical results with three trigger transverse momentum ($p_{\rm T}^Z$) ranges, $15<p_{\rm T}^Z<30$, $30<p_{\rm T}^Z<60$, and $p_{\rm T}^Z>60$ GeV, are compared with the ATLAS data \cite{ATLAS:2020wmg}, respectively. Numerical results describe the data well, especially at large $p_{\rm T}^{h}$. The theoretical calculations underestimate the data at low $p_{\rm T}^{h}$ because the fragmentation functions are limited with $z_{\min}=0.05$ and the more higher order corrections should be taken into account.

\begin{figure*}[!ht]
\centering
\includegraphics[width=1\linewidth]{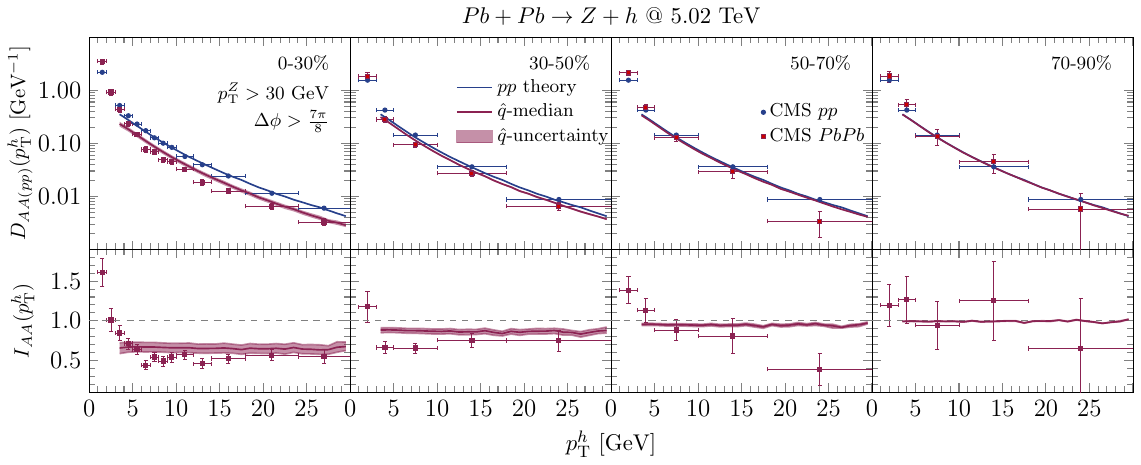}
\caption{(Color online) The hadron $p_{\rm T}^{h}$ distributions with the $Z$ trigger $p_{\rm T}^Z>30$ GeV in $pp$ (dashing curves) and 0-30\%, 30-50\%, 50-70\% and 70-90\% $PbPb$ (solid curves with bands) collisions at $\sqrt{s_{\rm NN}}=5.02$ TeV are shown in the upper panels, while the corresponding $I_{AA}(p_{\rm T}^h)$ in the lower panels, compared with the CMS experimental data \cite{CMS:2021otx}, respectively.}
\label{fig:CMS_pt}
\end{figure*}

\begin{figure*}[!ht]
\centering
\includegraphics[width=1\linewidth]{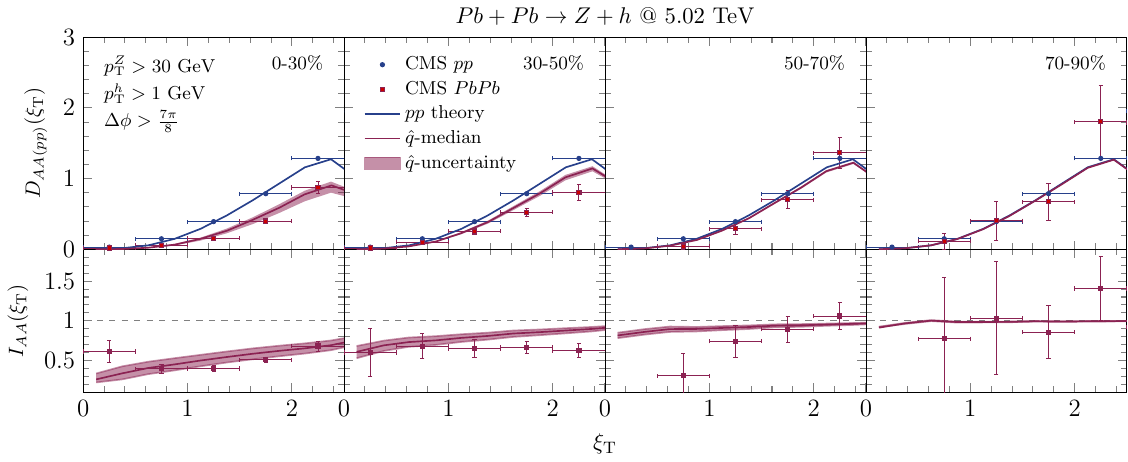}
\caption{(Color online) The $Z$-hadron $\xi_{\rm T}$ distributions with the $Z$ trigger $p_{\rm T}^Z>30$ GeV and the correlated haron $p_{\rm T}^h>1$ GeV in $pp$ (dashing curves) and 0-30\%, 30-50\%, 50-70\% and 70-90\% $PbPb$ (solid curves with bands) collisions at $\sqrt{s_{\rm NN}}=5.02$ TeV are shown in the upper panels, while the corresponding $I_{AA}(\xi_{\rm T})$ in the lower panels, compared with the CMS experimental data \cite{CMS:2021otx}, respectively.}
\label{fig:CMS_xi}
\end{figure*}

\begin{figure}[!ht]
    \centering
    \includegraphics[width=1\linewidth]{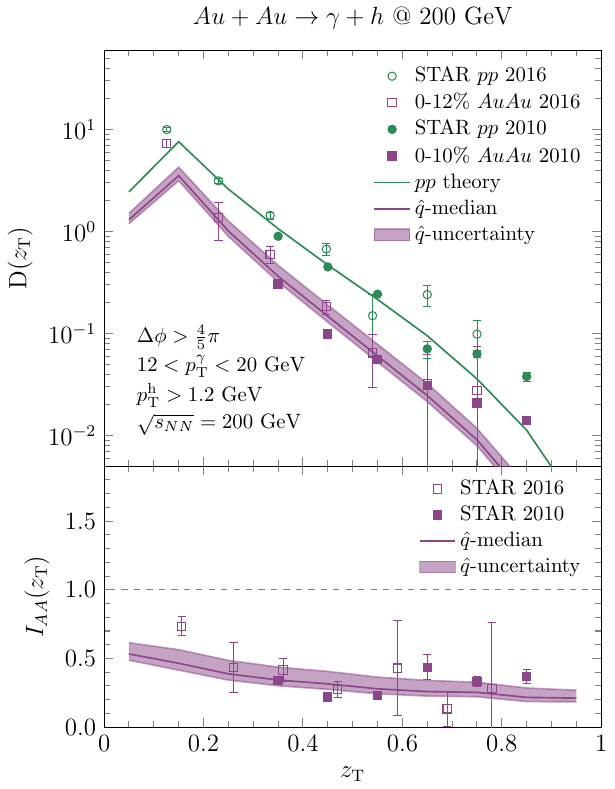}
    \caption{(Color online) The $\gamma$-hadron $z_{\rm T}$ distributions with the $\gamma$ trigger $12<p_{\rm T}^\gamma<20$ GeV and the correlated haron $p_{\rm T}^h>1.2$ GeV in $pp$ (dashing curve) and 0-10\% $AuAu$ (solid curve with band) collisions at $\sqrt{s_{\rm NN}}=200$ GeV are shown in the upper panel, while the corresponding $I_{AA}(z_{\rm T})$ in the lower panel, compared with the STAR experimental data \cite{STAR:2009ojv,STAR:2016jdz}, respectively.}
    \label{fig:STAR_zt}
\end{figure}

\begin{figure}[!ht]
\centering
\includegraphics[width=1\linewidth]{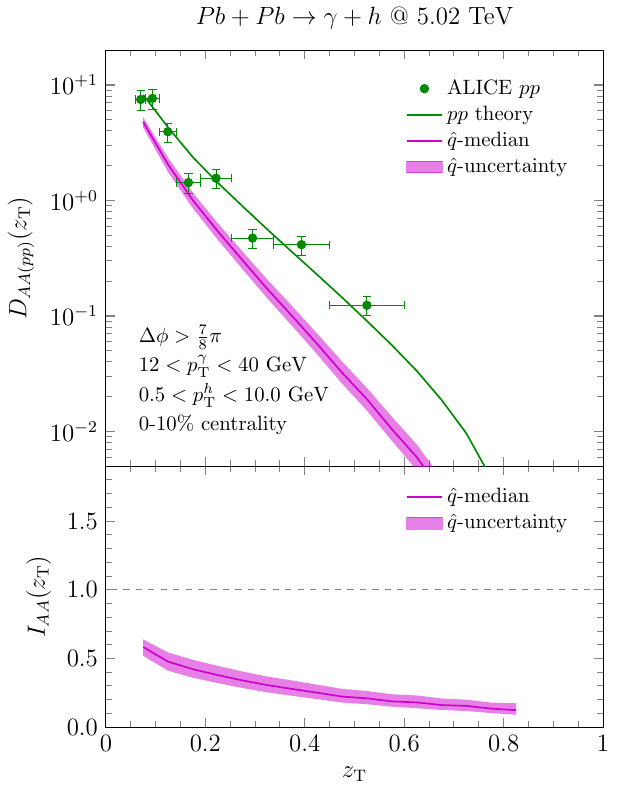}
\caption{(Color online) The $\gamma$-hadron $z_{\rm T}$ distributions with the $\gamma$ trigger $12<p_{\rm T}^\gamma<40$ GeV and the correlated haron $0.5<p_{\rm T}^h<10.0$ GeV in $pp$ (dashing curve) and 0-10\% $PbPb$ (solid curve with band) collisions at $\sqrt{s_{\rm NN}}=5.02$ TeV are shown in the upper panel, while the corresponding $I_{AA}(z_{\rm T})$ in the lower panel. The $pp$ spectra are compared with the ALICE experimental data \cite{ALICE:2020atx}.}
\label{fig:ALICE_zt}
\end{figure}

\begin{figure*}[!ht]
\centering
\includegraphics[width=0.8\linewidth]{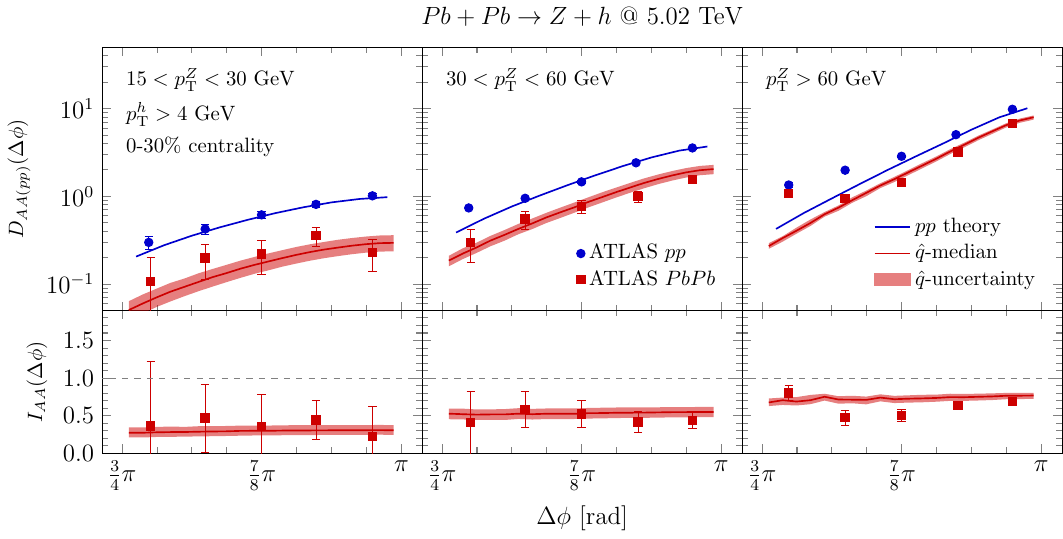}
\caption{(Color online) The per-trigger yield of $Z$-hadron pairs as a function of $\Delta {\phi}$ with the $Z$ trigger $15<p_{\rm T}^Z<30$, $30<p_{\rm T}^Z<60$ and $p_{\rm T}^Z>60$ GeV (from left to right panels) and the associated hadron $p_{\rm T}^h>4$ GeV in $pp$ (dashing curves) and 0-30\% $PbPb$ (solid curves) collisions at $\sqrt{s_{\rm NN}}=5.02$ TeV are shown in the upper panels, while the corresponding $I_{AA}(\Delta\phi)$ in the lower panels, compared with the ATLAS experimental data \cite{ATLAS:2020wmg}. }
\label{fig:ATLAS_dphi}
\end{figure*}

The nuclear modification factors $I_{AA}(p_{\rm T}^{h})$ of $Z$-hadrons in 0-10\% (upper panels) and 10-30\% (lower panels) $PbPb$ collisions at $\sqrt{s_{\rm NN}}=5.02$ TeV are shown in \autoref{fig:ATLAS_iaa}, compared with the ATLAS data, respectively. From the left to the right panel, the transverse momentum of trigger $Z$ is set as $15<p_{\rm T}^Z<30$ GeV, $30<p_{\rm T}^Z<60$ GeV and $p_{\rm T}^Z>60$ GeV. The azimuthal angle between the vector boson and the hadron is set as $\Delta\phi = |\phi_{V}-\phi_{h}|>3\pi/4$.
The solid curves are the numerical results with the median value of the $\hat{q}/T^3$, which is extracted from hadron suppression at RHIC and the LHC energies using the IF-Bayesian method \cite{Xie:2022ght,Xie:2022fak}, and the bands are the estimated uncertainty given by $\hat{q}/T^3$. It is worth mentioning that the suppression uncertainties shown in \autoref{fig:ATLAS_iaa} are contributed by the parton energy loss, which is because the medium-induced broadening mainly affects the azimuthal angular of jets and hardly influences their yield \cite{Wang:1998ww}. Our numerical results are in good agreement with the experimental data at mid to large $p_{\rm T}^{h}$ for the different trigger $p_{\rm T}^Z$.

No matter the experimental data or the numerical results in \autoref{fig:ATLAS_iaa}, show that the $Z$-hadron suppression becomes weak when the transverse momentum of trigger $Z$ increases.
This phenomenon is consistent with our previous studies on di-/$\gamma$-hadrons \cite{Zhang:2007ja,Zhang:2009rn,Xie:2019oxg}, by the difference in the production positions of jets with different momentum. The hadrons triggered by the large $p_{\rm T}^Z$ $Z$ bosons are from the fragmentation of the hard partons which are more likely to be produced at the surface of the medium and prefer to emit off the medium perpendicularly with less energy loss, while the hadrons triggered by the small $p_{\rm T}^Z$ $Z$ bosons are contributed by the small $p_{\rm T}$ jets which are produced at the center of the medium and emit off the medium with large energy loss fractions.
Numerical results underestimate the suppression with the large trigger $p_{\rm T}^Z>60$ GeV, which is similar to the previous studies on $\gamma$-hadron \cite{Xie:2020zdb} and dihadron \cite{Xie:2019oxg}. It might be because the parton energy dependence was not taken into account in the IF-Bayesian $\hat{q}$ extraction method \cite{Xie:2022ght,Xie:2022fak}, and the energy loss fraction $\Delta{E}/E$ decrease with the parton energy increasing.

To further study the centrality dependence of $Z$-hadron transverse momentum correlations, we show in \autoref{fig:CMS_pt} for the $Z$-triggered hadron spectra $D_{AA}(p_{\rm T}^h)$ (upper panels) and the corresponding suppression factors $I_{AA}(p_{\rm T}^h)$ (lower panels) with the trigger $p_{\rm T}^Z>30$ GeV in 0-30\%, 30-50\%, 50-70\% and 70-90\% $PbPb$ collisions at 5.02 TeV, respectively. The azimuthal angle between the $Z$ boson and the hadron is set as $\Delta\phi>7\pi/8$. The spectra $D_{pp}(p_{\rm T}^h)$ in $pp$ collisions at 5.02 TeV are also shown in the upper panels.
We can see the numerical results fit CMS data \cite{CMS:2021otx} well.
In most central collisions, the $Z$-triggered hadrons could be suppressed by about 30-50\% due to jet energy loss. Such suppression becomes weaker and weaker until jet energy loss almost vanishes from central to peripheral collisions since the medium size becomes small in peripheral collisions. Meanwhile, the theoretical uncertainty related to jet energy loss also becomes small in peripheral collisions.

\begin{figure}[!ht]
\centering
\includegraphics[width=1\linewidth]{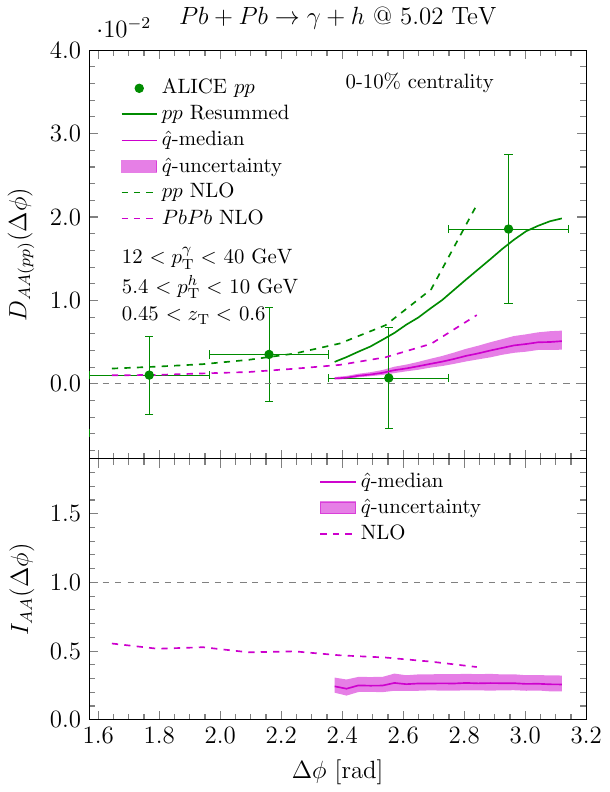}
\caption{(Color online) The per-trigger yield of $\gamma$-hadron pairs as a function of $\Delta {\phi}$ with the $\gamma$ trigger $12<p_{\rm T}^{\gamma}<40$ GeV and the associated hadron $5.4<p_{\rm T}^h<10$ GeV or $0.45<z_{\rm T}<0.6$ in $pp$ (green curves) and 0-10\% $PbPb$ (pink curves) collisions at $\sqrt{s_{\rm NN}}=5.02$ TeV are shown in the upper panel, while the corresponding $I_{AA}(\Delta\phi)$ in the lower panel. The solid (dashing) curves are calculated with the resummation-improved (the $\alpha_s^3$ fixed-order or NLO) pQCD parton model for $pp$ and $AA$ collisions, respectively. The $D_{pp}(\Delta {\phi})$ data are from the ALICE experiments \cite{ALICE:2020atx}.}
\label{fig:ALICE_dphi}
\end{figure}

\begin{figure}[!ht]
\centering
\includegraphics[width=1\linewidth]{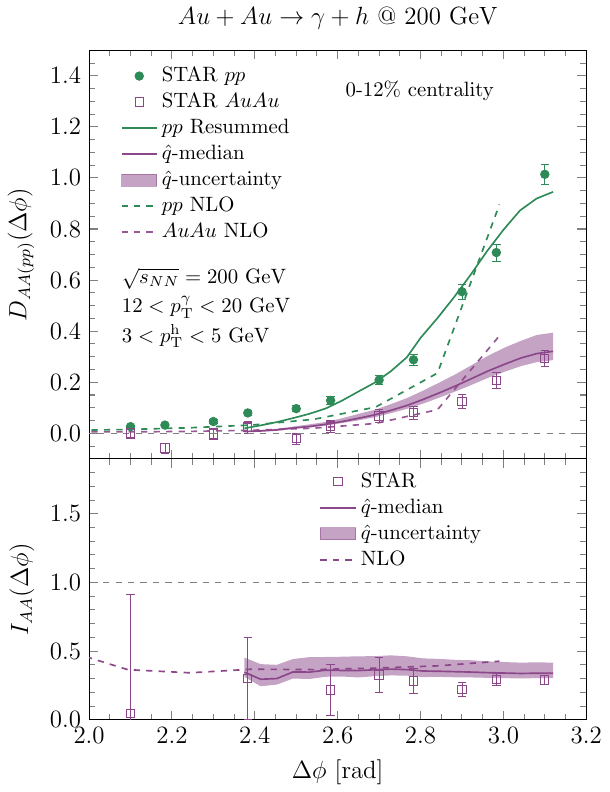}
\caption{(Color online) The per-trigger yield of $\gamma$-hadron pairs as a function of $\Delta {\phi}$ with the $\gamma$ trigger $12<p_{\rm T}^{\gamma}<20$ GeV and the associated hadron $3<p_{\rm T}^h<5$ GeV in $pp$ (green curves) and 0-12\% $AuAu$ (pink curves) collisions at $\sqrt{s_{\rm NN}}=200$ GeV are shown in the upper panel, while the corresponding $I_{AA}(\Delta\phi)$ in the lower panel, compared with the STAR experimental data \cite{STAR:2016jdz}. The solid (dashing) curves are calculated with the resummation-improved (the $\alpha_s^3$ fixed-order or NLO) pQCD parton model for $pp$ and $AA$ collisions, respectively.}
\label{fig:STAR_dphi}
\end{figure}

Shown in \autoref{fig:CMS_xi} are the $Z$-hadron $\xi_{\rm T}$ distributions $D_{AA}(\xi_{\rm T})$ (upper panels) and the corresponding suppression factors $I_{AA}(\xi_{\rm T})$ (lower panels) with $Z$ trigger $p_{\rm T}^Z>30$ GeV and the away-side associated haron $p_{\rm T}^h>1$ GeV in 0-30\%, 30-50\%, 50-70\% and 70-90\% $PbPb$ collisions at 5.02 TeV, respectively. The azimuthal angle between the $Z$ boson and the hadron is set as $\Delta\phi>7\pi/8$. The spectra $D_{pp}(\xi_{\rm T})$ in $pp$ collisions at 5.02 TeV are also shown in the upper panels.
According to the definitions of $\xi_{\rm T}=\ln (-\left|\vec{p}_{\rm T}^{Z}\right|^2 /\vec{p}_{\rm T}^{h} \cdot \vec{p}_{\rm T}^{Z}) \approx -\ln z_{\rm T}$ and $z_{\rm T}=p_{\rm T}^{h}/p_{\rm T}^Z$, $\xi_{\rm T}$ can be understood as the magnitude of the unit vector where the hadron transverse momentum is projected on the trigger axis, while the negative sign is for the projection in the away-side hemisphere. We note that a hadron carrying all of the parent parton momentum, $z_{\rm T}=1$, and moving back-to-back away from the trigger, $\Delta\phi=\pi$, would have a $\xi_{\rm T}=0$.
Because the calculations for low $p_{\rm T}^{h}$ hadrons are limited by the fragmentation limit $z_{\min}=0.05$, we focus on the region of $\xi_{\rm T}\leq2.5$, which corresponds to $z_{\rm T}\geq0.08$.
An indication of energy loss would mean that the $Z$-hadron spectrum of $PbPb$ collisions has a shift to a larger value of $\xi_{\rm T}$ compared to that in $pp$ collisions, which is more evident when comparing the results from peripheral collisions to central collisions.
Our calculations can provide a good description to the CMS measurements for $Z$-hadron $\xi_{\rm T}$ distributions and the supperresions $I_{AA}(\xi_{\rm T})$, especially for most central $AA$ collisions.

Similar calculations were done for $\gamma$-triggered hadrons, as shown in \autoref{fig:STAR_zt} with the $\gamma$ trigger $12<p_{\rm T}^\gamma<20$ GeV and the correlated haron $p_{\rm T}^h>1.2$ GeV in $pp$ and 0-10\% $AuAu$ collisions at $\sqrt{s_{\rm NN}}=200$ GeV, and in \autoref{fig:ALICE_zt} with the $\gamma$ trigger $12<p_{\rm T}^\gamma<40$ GeV and the correlated haron $0.5<p_{\rm T}^h<10.0$ GeV in $pp$ and 0-10\% $PbPb$ collisions at $\sqrt{s_{\rm NN}}=5.02$ TeV. The jet energy loss parameter $\hat{q}$ and its uncertainty are all set the same as in \autoref{fig:ATLAS_iaa}, \autoref{fig:CMS_pt}, and \autoref{fig:CMS_xi}. Notice that the IF-Bayesian extracted $\hat{q}$ is available for different-centrality $AA$ collisions at RHIC and the LHC \cite{Xie:2022ght,Xie:2022fak}. 
Our numerical results for the $\gamma$-triggered fragmentation functions $D(z_{\rm T})$ and the nuclear modification factor $I_{AA}(z_{\rm T})$ both describe the STAR experimental data \cite{STAR:2009ojv,STAR:2016jdz} very well, which demonstrates the jet quenching description and the Sudakov resummation improved pQCD model are effective no matter at the LHC energies or RHIC energies. Besides for the well fitting to ALICE data  \cite{ALICE:2020atx} for $pp$ collisions in  \autoref{fig:ALICE_zt}, we predict the $\gamma$-triggered away-side hadron spectra $D_{AA}(z_{\rm T})$ and the suppression factor $I_{AA}(z_{\rm T})$ in 0-10$\%$ $PbPb$ collisions at $\sqrt{s_\mathrm{NN}} = 5.02$ TeV.
Numerical results show that $\gamma$-triggered hadron spectra are suppressed by 40\% for low $z_{\rm T}$ and 80\% for large $z_{\rm T}$ due to jet energy loss in central $PbPb$ collisions.

We now have a well-established pQCD framework with the Sudakov resummation technique, which can well describe the transverse momentum correlations of $Z/\gamma$-hadron pairs via comparisons with the experimental data in $pp$ and $AA$ collisions at both RHIC and LHC. 
Next let's check $\left\langle p_{\perp}^2 \right\rangle$ as well as $\Delta E$ effects on the azimuthal angular ($\Delta \phi$) correlations of $Z/\gamma$-hadron production in central $AA$ collisions at RHIC and the LHC. 

\subsection{azimuthal angular correlations of \texorpdfstring{$Z/\gamma$}{Z/gamma}-hadrons}

Shown in the upper panels of \autoref{fig:ATLAS_dphi} are the per-trigger yield of $Z$-hadron pairs as a function of the azimuthal angle $\Delta {\phi}$ between the $Z$ trigger (with $15<p_{\rm T}^Z<30$, $30<p_{\rm T}^Z<60$ and $p_{\rm T}^Z>60$ GeV from left to right panels) and the associated hadron (with $p_{\rm T}^h>4$ GeV) in $pp$ (dashing curves) and 0-30\% $PbPb$ (solid curves) collisions at $\sqrt{s_{\rm NN}}=5.02$ TeV, and in the lower panels are the corresponding nuclear modificatio factors $I_{AA}(\Delta\phi)$. Our theoretical calculations can provide a very good description to the ATLAS experimental data \cite{ATLAS:2020wmg} in both $pp$ and $PbPb$ collisions for three trigger $p_{\rm T}^Z$ ranges.
Numerical results show that with the trigger $p_{\rm T}^Z$ increasing, the angular correlation of $Z$-hadron pairs becomes stronger in both $pp$ and central $AA$ collisions but the yield suppression changes weaker. The former is mainly because the Sudakov effect enhances as the $Q^2$ increases, as shown in Eq. (\ref{eq:sud_p}), and the latter is similar as in \autoref{fig:ATLAS_iaa}. 
The uncertainty from the medium-induced broadening is almost swallowed up by that from the jet energy loss. The approximate-constant $I_{AA}(\Delta\phi)$ as a function of $\Delta\phi$, as shown in the lower panels of \autoref{fig:ATLAS_dphi}, means the medium-induced broadening effect is fairly weak. This is consistent with previous studies on dijet angular correlations due to the dominant Sudakov effect for the large $p_T$ jets, especially at LHC energies  \cite{Chen:2016vem,Chen:2016cof}.

In \autoref{fig:ALICE_dphi} we present the per-trigger yield of $\gamma$-hadron pairs as a function of $\Delta {\phi}$ with the $\gamma$ trigger $12<p_{\rm T}^{\gamma}<40$ GeV and the associated hadron $5.4<p_{\rm T}^h<10$ GeV or $0.45<z_{\rm T}<0.6$ in $pp$ and 0-10\% $PbPb$ collisions at $\sqrt{s_{\rm NN}}=5.02$ TeV in the upper panel, while the corresponding $I_{AA}(\Delta\phi)$ in the lower panel.
Here we use the Sudakov resummation-improved pQCD parton model to focus on the back-to-back region (in fact we present in the region $2.2<\Delta {\phi}<\pi$) in which the $\alpha_s^3$ fixed-order or NLO pQCD expansion for $D_{pp(AA)}(\Delta {\phi})$ diverges. For a comparison, we also give the NLO results  \cite{Xie:2022ght,Xie:2022fak} in the region $1.65<\Delta {\phi}<2.85$.
In fitting to the ALICE data \cite{ALICE:2020atx} for $D_{pp}(\Delta {\phi})$, the two models both work well in their respective $\Delta {\phi}$ regions. 
We also predict the $\gamma$-triggered away-side hadron spectra $D_{AA}(\Delta {\phi})$ and the suppression factor $I_{AA}(\Delta {\phi})$ in 0-10$\%$ $PbPb$ collisions at $\sqrt{s_\mathrm{NN}} = 5.02$ TeV. Compared to the $\Delta \phi$ distribution of $Z$-hadron production in \autoref{fig:ALICE_zt} with the same kinematic cuts, the distribution of $\gamma$-hadron gets almost the same suppression about $I_{AA}\approx$ 0.25 in \autoref{fig:ALICE_dphi} when both energy loss and the medium-induced broadening with median $\hat{q}/T^3$ are included in the calculations for $PbPb$ collisions as opposed to that for $pp$ collisions.

\begin{figure*}[!ht]
\centering
\includegraphics[width=1\linewidth]{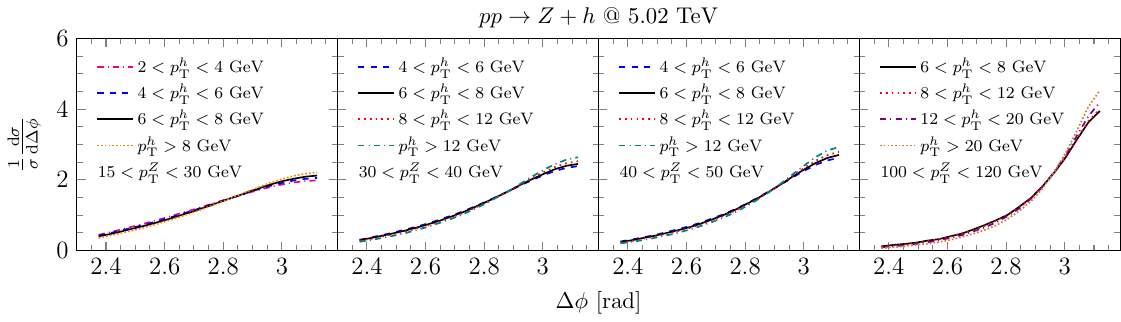}
\caption{(Color online) Normalized $Z$-hadron angular distributions in $pp$ collisions at $\sqrt{s_{\rm NN}}=5.02$ TeV with a fixed trigger-$p_{\rm T}^Z$ region and different association $p_{\rm T}^h$ regions.
The four fixed trigger-$p_{\rm T}^Z$ regions are $15<p_{\rm T}^Z<30$, $30<p_{\rm T}^Z<40$, $40<p_{\rm T}^Z<50$ and $100<p_{\rm T}^Z<120$ GeV from the left to the right panel, respectively.}
\label{fig:ATLAS_NormDphi_pp}
\end{figure*}

\begin{figure*}[!ht]
\centering
\includegraphics[width=1\linewidth]{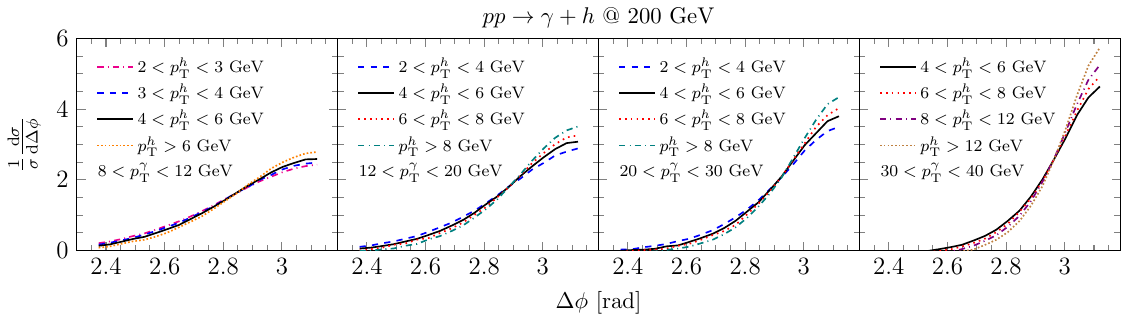}
\caption{(Color online) Similar to \autoref{fig:ATLAS_NormDphi_pp} but for the $\gamma$-hadron production at $\sqrt{s_{\rm NN}}=200$ GeV. The kinematics cuts are different correspondingly.}
\label{fig:STAR_DphiNorm_pp}
\end{figure*}


Shown in \autoref{fig:STAR_dphi} is the $\Delta \phi$ distribution of the $\gamma$-hadron correlation with the trigger $12<p_{\rm T}^{\gamma}<20$ GeV and the association $3<p_{\rm T}^h<5$ GeV in $pp$ and 0-12\% $AuAu$ collisions at $\sqrt{s_{\rm NN}}=200$ GeV in the upper panel, while the suppression factor in the lower panel, compared with the STAR measurements \cite{STAR:2016jdz}, respectively. 
The numerical results from the resummation-improved pQCD parton model provide a very well fit to the experimental data for $D_{pp}(\Delta {\phi})$, $D_{AA}(\Delta {\phi})$ and $I_{AA}(\Delta {\phi})$ in the region of $2.4<\Delta {\phi}<\pi$, respectively, while the NLO calculations fit well in $1.6 (2.0) <\Delta {\phi}<2.4$.

\subsection{RMS analysis of the width of \texorpdfstring{$\Delta \phi$}{dphi} distribution}

\begin{figure}[!ht]
\centering
\includegraphics[width=1\linewidth]{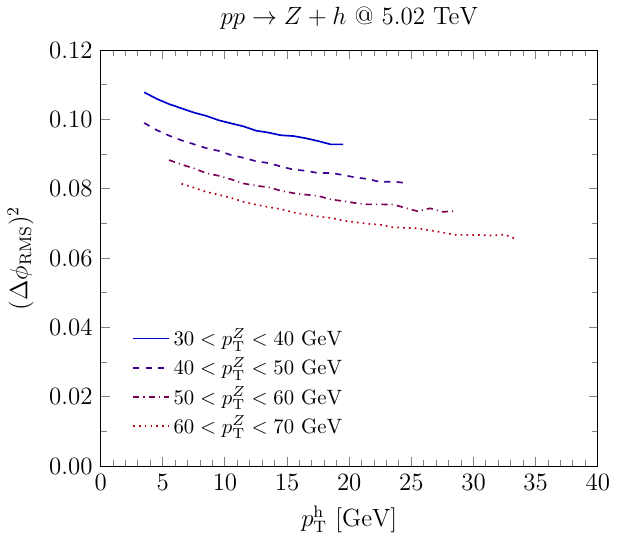}
\caption{(Color online)  RMS width of angular distributions of $Z$-hadron as a function of $p_{\rm T}^h$ in $pp$ collisions at $\sqrt{s_{\rm NN}}=5.02$ TeV with four different $p_{\rm T}^Z$ regions, $30<p_{\rm T}^Z<40$, $40<p_{\rm T}^Z<50$, $50<p_{\rm T}^Z<60$ and $60<p_{\rm T}^Z<70$ GeV from top to bottom, respectively.}
\label{fig:ATLAS_DphiRMS_pth_pp}
\end{figure}

\begin{figure}[!ht]
\centering
\includegraphics[width=1\linewidth]{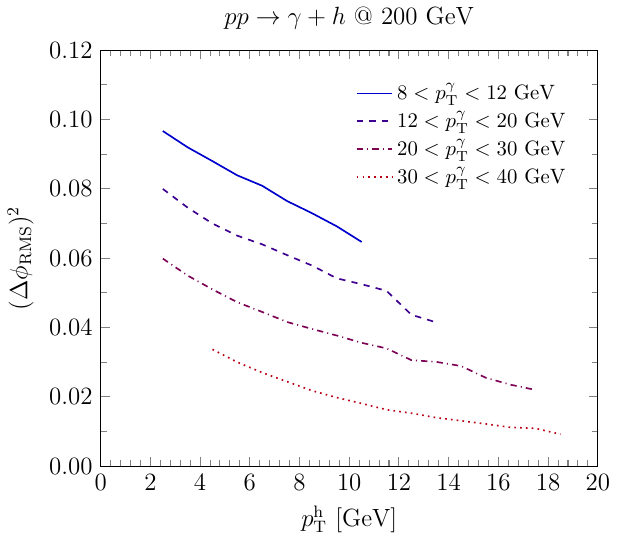}
\caption{(Color online) Similar to \autoref{fig:ATLAS_DphiRMS_pth_pp} but for the $\gamma$-hadron production at $\sqrt{s_{\rm NN}}=200$ GeV. The kinematics cuts are different correspondingly.}
\label{fig:STAR_DphiRMS_pth_pp}
\end{figure}

In the above calculations the Sudakov resummation technique is introduced into the pQCD parton model to get rid of the divergence on the endpoints in the $\Delta \phi$ distribution of $Z/\gamma$-hadron pairs in $pp/AA$ collisions. The numerical results show that the Sudakov resummation improved pQCD parton model works well not only on the transverse momentum correlation but also on the azimuthal angular correlation when $\Delta \phi$ is constrained near the back-to-back region of $Z/\gamma$-hadron pairs.
Especially, after including the jet energy loss $\Delta E$ and the medium-induced broadening $\langle p_{\perp}^2 \rangle$ in high-energy $AA$ collisions, the model succeeds in describing the suppression of the large-$p_{\rm T}$ $Z/\gamma$-hadron as well as the medium-induced broadening effect.
In the $Z/ \gamma$-hadron correlations, the strength of the azimuthal angular correlation is of primary concern. In $pp$ collisions, the $\Delta \phi$ correlation is contributed by the vacuum Sudakov effect. While in $AA$ collisions, a medium-induced $\langle p_{\perp}^2 \rangle$ effect is included (see Eq.({\ref{eq:sud_med}})) in the vacuum Sudakov factor  and the jet energy loss $\Delta E$ is adopted (see Eq.{(\ref{eq:nff}})) to modify the fragmentation function. 
In this subsection, we further analyze the angular correlation by normalizing the $\Delta \phi$ distribution to unity. This approach allows us to focus on the differences between $pp$ and $AA$ collisions with $\Delta E$ or $\langle p_{\perp}^2 \rangle$, or both.

Shown in each panel of \autoref{fig:ATLAS_NormDphi_pp} are the normalized $Z$-hadron angular distributions in $pp$ collisions at $\sqrt{s_{\rm NN}}=5.02$ TeV with a fixed trigger-$p_{\rm T}^Z$ region and different association $p_{\rm T}^h$ regions.
The four fixed trigger-$p_{\rm T}^Z$ regions are $15<p_{\rm T}^Z<30$, $30<p_{\rm T}^Z<40$, $40<p_{\rm T}^Z<50$ and $100<p_{\rm T}^Z<120$ GeV from the left to the right panel, respectively. \autoref{fig:STAR_DphiNorm_pp} is similar to \autoref{fig:ATLAS_NormDphi_pp} but for $\gamma$-hadron production at $\sqrt{s_{\rm NN}}=200$ GeV with different kinematics cuts correspondingly. The numerical results at both LHC and RHIC show that the angular correlation is stronger with the $p_{\rm T}$ increasing of both the trigger $Z/\gamma$ and the associated hadron. With current selections of $p_{\rm T}$ cuts, the $p_{\rm T}$ dependence of the angular correlation is weaker at LHC than at RHIC.

The angular correlation strength can be quantified by the root-mean-square (RMS) width of $\Delta \phi$ distribution \cite{Jia:2019qbl}:
\begin{equation}
\Delta \phi_{\mathrm{RMS}}=\sqrt{\frac{\int \dd \Delta \phi(\Delta \phi-\pi)^2 \frac{\dd \sigma}{\dd \Delta \phi}}{\int \dd \Delta \phi \frac{\dd \sigma}{\dd \Delta \phi}}},
\end{equation}
where the integral range of $\Delta \phi$ is chosen from 2.4 to $\pi$ to constrain $Z/\gamma$-hadron pairs near the back-to-back region. For a totally back-to-back configuration, corresponding to a delta function for the azimuthal distribution, $\Delta\phi_{\rm RMS}=0$, which gives the strongest angular correlation. The larger value of $\Delta\phi_{\rm RMS}$ would indicate a broader azimuthal distribution or a weaker angular correlation.

Shown in \autoref{fig:ATLAS_DphiRMS_pth_pp} (\autoref{fig:STAR_DphiRMS_pth_pp}) is 
the RMS width of angular distributions of $Z$-hadron ($\gamma$-hadron) as a function of $p_{\rm T}^h$ in $pp$ collisions at $\sqrt{s_{\rm NN}}=5.02$ TeV (200 GeV) with four different $p_{\rm T}^Z$ ($p_{\rm T}^{\gamma}$) regions, respectively. The conclusion is the same as that in \autoref{fig:ATLAS_NormDphi_pp} and \autoref{fig:STAR_DphiNorm_pp}. 
The RMS width is dependent on the transverse momentum of both the trigger and the associated hadron, and the dependence on $p_{\rm T}^Z$ or $p_{\rm T}^\gamma$ is slightly stronger than $p_{\rm T}^h$. In following calculations we will choose a fixed region for $p_{\rm T}^h$ with varying $p_{\rm T}^Z$ or $p_{\rm T}^\gamma$.

\begin{figure*}[!ht]
\centering
\subcaptionbox*{}{\includegraphics[width=0.32\linewidth]{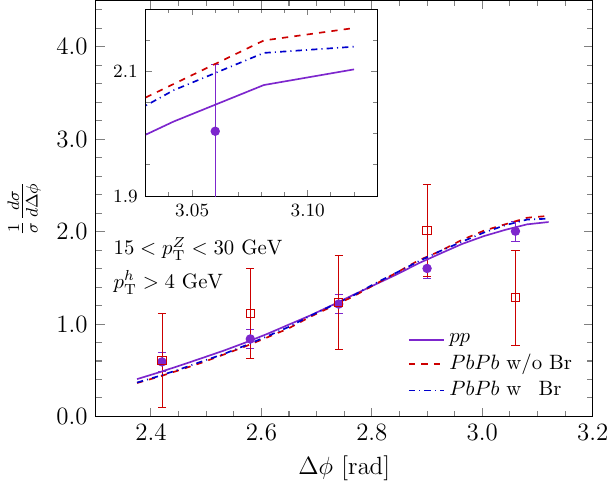}}
\hfill
\subcaptionbox*{}{\includegraphics[width=0.32\linewidth]{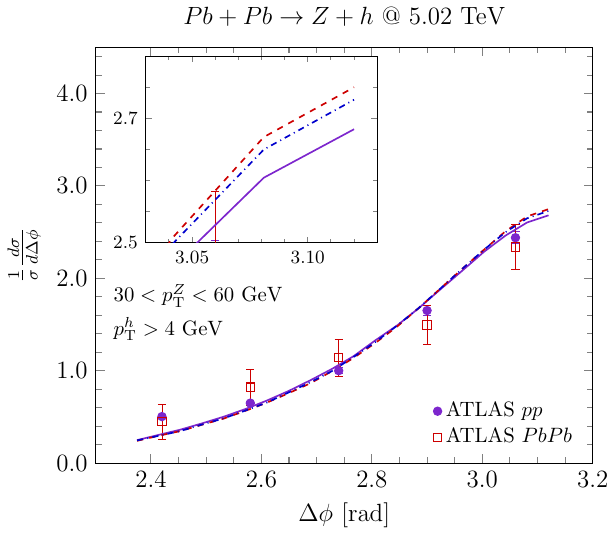}}
\hfill
\subcaptionbox*{}{\includegraphics[width=0.32\linewidth]{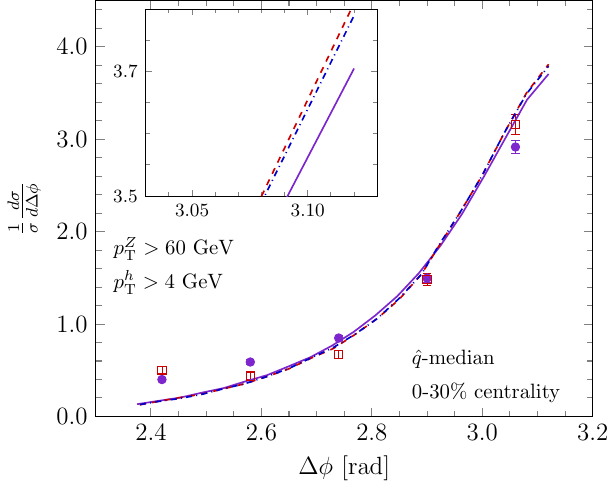}}
\hfill
\caption{(Color online) Normalized $Z$-hadron angular distributions in $pp$ (solid purple) collisions and 0-10\% $PbPb$ collisions at $\sqrt{s_{\rm NN}}=5.02$ TeV with (blue dot-dashed) and without (red dashed) the medium-induced broadening effect. 
Different $p_{\rm T}$ cuts are set for the triggers and the correlated hadrons in the panels, respectively. The data are from the ATLAS experiment \cite{ATLAS:2020wmg}.
}
\label{fig:ATLAS_NormDphi_ppPbPb}
\end{figure*}

\begin{figure*}[!ht]
\centering
\subcaptionbox*{}{\includegraphics[width=0.32\linewidth]{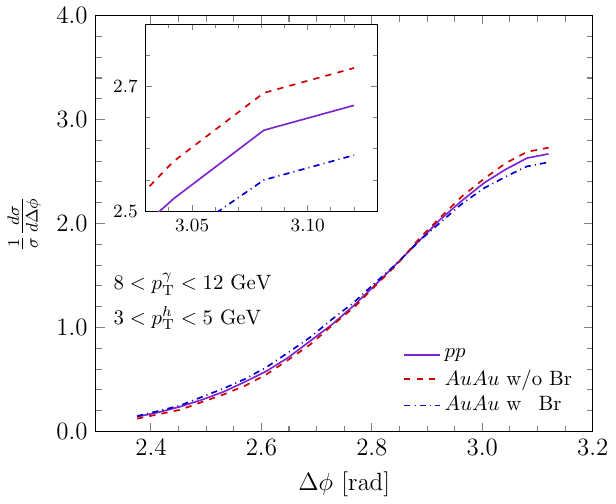}}
\hfill
\subcaptionbox*{}{\includegraphics[width=0.32\linewidth]{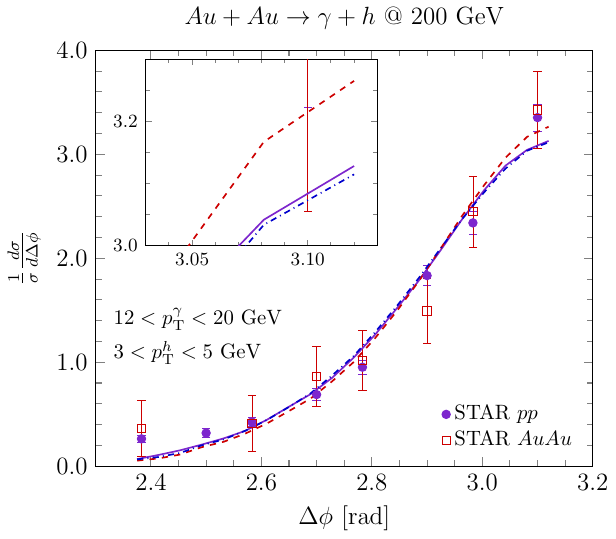}}
\hfill
\subcaptionbox*{}{\includegraphics[width=0.32\linewidth]{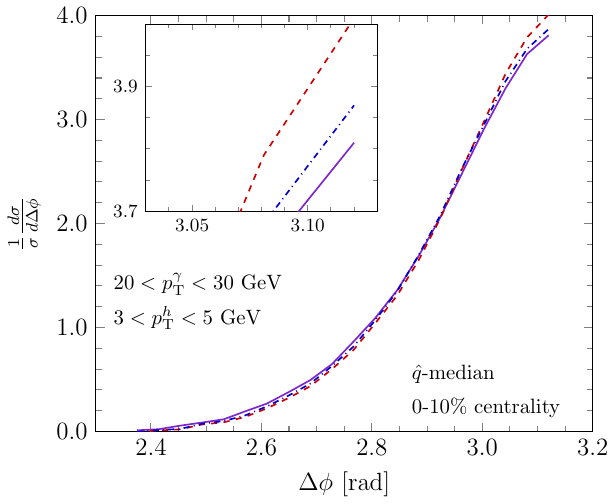}}
\hfill
\caption{(Color online) Similar to \autoref{fig:ATLAS_NormDphi_ppPbPb} but for the $\gamma$-hadron production in 0-10\% $AuAu$ collisions at $\sqrt{s_{\rm NN}}=200$ GeV. The kinematics cuts are different correspondingly. The data are from the STAR experiment \cite{STAR:2016jdz}.}
\label{fig:STAR_NormDphi_ppAuAu}
\end{figure*}

To see the influence of jet quenching effect on the azimuth angular correlation, we add the energy loss and transverse momentum broadening into $AA$ collisions progressively and compare the $\Delta \phi$ distribution with $pp$ results.
Shown in \autoref{fig:ATLAS_NormDphi_ppPbPb} are the normalized $Z$-hadron angular distributions in $pp$ and 0-10\% $PbPb$ collisions at $\sqrt{s_{\rm NN}}=5.02$ TeV with only the energy loss and both the energy loss and the medium-induced broadening effect. The inserted panels are to show the results near $\pi$ clearly. Different $p_{\rm T}$ cuts are set for the triggers and the correlated hadrons in the different panels, respectively.
\autoref{fig:STAR_NormDphi_ppAuAu} is similar to \autoref{fig:ATLAS_NormDphi_ppPbPb} but for the $\gamma$-hadron production in 0-10\% $AuAu$ collisions at $\sqrt{s_{\rm NN}}=200$ GeV. The kinematics cuts are different correspondingly. 
Here only the median value of $\hat{q}$ from the IF-Bayesian extraction is used for the energy loss and and the medium-induced broadening in central $AA$ collisions at both RHIC and LHC. 
First of all, the intuitive overall phenomenon is similar to the $pp$ cases, even in central $AA$ collisions at both RHIC and LHC the angular correlation is stronger with the $p_{\rm T}$ increasing of both the trigger $Z/\gamma$ and the associated hadron.

For central $PbPb$ (\autoref{fig:ATLAS_NormDphi_ppPbPb}) 
and $AuAu$ (\autoref{fig:STAR_NormDphi_ppAuAu}) 
collisions, we set two cases, one is only with the energy loss effect but without the medium-induced broadening effect (red dashed curves), and the other is with both of them (blue dot-dashed curves). 
When only considering the jet energy loss effect in $AA$ collisions, one can observe that the angular correlation is stronger in $AA$ collisions than that in $pp$ collisions (blue solid curves). After taken with both the energy loss $\Delta{E}$ and the medium-induced broadening $\langle p_{\perp}^2 \rangle$ in $AA$ collisions, the normalized $\Delta \phi$ distributions are then broadened. Compared with $pp$ collisions, in the first panels of Figure \ref{fig:STAR_NormDphi_ppAuAu} the normalized $\Delta \phi$ distributions in $AA$ collisions are broader than in $pp$ collisions. However, in the other panels of Figure \ref{fig:STAR_NormDphi_ppAuAu} and Figure \ref{fig:ATLAS_NormDphi_ppPbPb}, despite the medium-induced broadening effect makes the normalized $\Delta \phi$ distributions broader, the normalized $\Delta \phi$ distributions in $AA$ collisions are narrower than in $pp$ collisions.
In a word, $\langle p_{\perp}^2 \rangle$ makes a decorrelation to broaden the $\Delta \phi$ distribution, while $\Delta{E}$ enhances the correlation to narrow the $\Delta \phi$ distribution, namely anti-broadening. The modification of the angular correlation is a result of the competition between the broadening and the anti-broadening. However, the medium modification of the sum of positive and negative contributions is too small to be observed noticeably via the comparison with pp and AA data due to large error bars.

Our numerical results fit data very well in Figure \ref{fig:STAR_NormDphi_ppAuAu} and Figure \ref{fig:ATLAS_NormDphi_ppPbPb}. We do not observe the evident broadening effect in $Z/\gamma$-hadron angular distributions in $AA$ collisions at both RHIC and LHC, which is different from the observation of the angular decorrelation in di-hadron/jet production \cite{Chen:2016vem,Chen:2016cof}. This is because of the weaker medium broadening effect in $Z/\gamma$-hadron production. The average broadening strength in one $Z/\gamma$-hadron/jet event is about half of that in one di-hadron/jet event due to the absence of the strong interaction between the $Z/\gamma$ boson and the medium. In addition, the fragmentation contribution ratio of hard quarks over gluons for the $Z/\gamma$-hadron production is larger than for the di-hadron production, and the $\langle p_{\perp}^2 \rangle$ value for a quark jet is $4/9$ times of that for a gluon jet.

To demonstrate the medium modifications, generally, one need to set the same kinematics cuts for $Z$/$\gamma$-hadron production between $AA$ and $pp$ collisions. Due to jet quenching or ``$p_{\rm T}$ shift", the parent parton for the observed hadrons in $AA$ collision has a larger $p_{\rm T}$ prior to quenching than the parent parton $p_{\rm T}$ in $pp$ collision. The more larger the $p_{\rm T}$ is, the more stronger the angular correlation is from the Sudakov effects, so the $Z$/$\gamma$-hadron angular correlation is enhanced by jet energy loss in $AA$ collisions due to the bias of the same kinematics cuts between $AA$ and $pp$ collisions. This is why the anti-broadening is brought by $\Delta{E}$. Numerical results in \autoref{fig:ATLAS_NormDphi_ppPbPb} and \autoref{fig:STAR_NormDphi_ppAuAu} show that the positive effect of the $\langle p_{\perp}^2 \rangle$ broadening can be either stronger or smaller than the negative effect of the $\Delta{E}$ anti-broadening in the angular correlations. The observed broadening displayed in the $\Delta \phi$ distribution is a sum of the positive and negative contributions. Such a competition 
between the broadening and the anti-broadening is stronger at RHIC than at LHC.

\begin{figure}[!ht]
\centering
\includegraphics[width=1\linewidth]{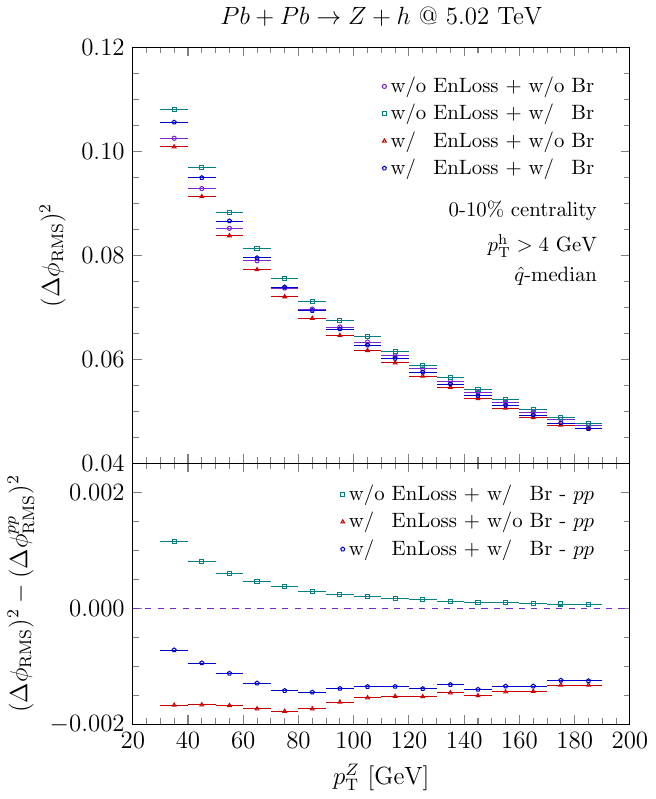}
\caption{(Color online) Shown in the upper panel are the RMS widths as a function of the trigger $p_{\rm T}^Z$ with the associated-hadron $p_{\rm T}>4$ GeV for the $Z$-hadron production in both $pp$ and central $PbPb$ collisions at 5.02 TeV. Either energy loss or medium broadening or both are set in $PbPb$ collisions. Their differences of the RMS widths against $pp$ collisions are shown in the lower panel.}
\label{fig:ATLAS_DphiRMS_ptZ}
\end{figure}

\begin{figure}[!ht]
\centering
\includegraphics[width=1\linewidth]{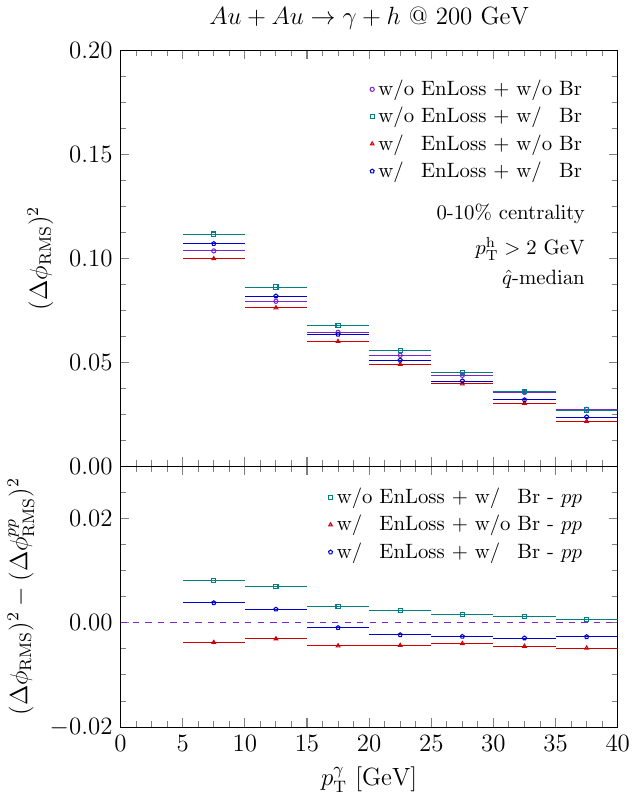}
\caption{(Color online) Similar to \autoref{fig:ATLAS_DphiRMS_ptZ} but as a function of $p_{\rm T}^{\gamma}$ for the $\gamma$-hadron production in $pp$ and central $AuAu$ collisions at 200 GeV.}
\label{fig:ATLAS_DphiRMS_ptGam}
\end{figure}

To quantify the broadening and the anti-broadening effects, we then plotted in the upper panel of \autoref{fig:ATLAS_DphiRMS_ptZ} the RMS width as a function of the trigger $p_{\rm T}^Z$ with the associated-hadron $p_{\rm T}>4$ GeV for the $Z$-hadron production in both $pp$ and central $PbPb$ collisions at 5.02 GeV. 
Three cases are set for the medium effects in $PbPb$ collisions, only with broadening effect (green square), only with energy loss effect (red triangle), and with both energy loss and broadening effects (blue pentagon), while the $pp$ results are given by purple circle.
Here we also use the $\hat{q}$-median for the jet energy loss and the medium-induced broadening.
We see that as we move to the trigger with the larger $p_{\rm T}^Z$ ($p_{\rm T}^{\gamma}$), the $\Delta\phi_{\rm RMS}$ decreases, corresponding to the sharper angular distribution or the stronger angular correlation.
Because both the energy loss and the medium broadening effects have weak effects at the LHC kinematics, it is difficult to differentiate them, we thus plot their difference against $pp$ calculations (without both effects) in the lower panel of  \autoref{fig:ATLAS_DphiRMS_ptZ}. 
\autoref{fig:ATLAS_DphiRMS_ptGam} is similar to \autoref{fig:ATLAS_DphiRMS_ptZ} but as a function of $p_{\rm T}^{\gamma}$ for the $\gamma$-hadron production in $pp$ and central $AuAu$ collisions at 200 GeV.
The medium-induced broadening has a positive azimuthal broadening effect on the RMS width, as shown (green square) in the both lower panels of \autoref{fig:ATLAS_DphiRMS_ptZ} and 
\autoref{fig:ATLAS_DphiRMS_ptGam}, 
which decreases as moving to large  $p_{\rm T}^{Z}$ or $p_{\rm T}^{\gamma}$ due to the overwhelming contribution from vacuum Sudakov effects.
The energy loss effect (red triangle) has a negative azimuthal broadening effect, so-called anti-broadening, and decreases slightly with $p_{\rm T}^{Z}$ or $p_{\rm T}^{\gamma}$. This is because the parton energy loss in the HT energy loss model has a logarithmic energy dependence that is weaker than a linear dependence, so the fractional energy loss $\Delta E/E$ decreases with jet energy \cite{Wang:2016fds}. 

The observed width difference should be the combined broadening effect (blue pentagon) from the medium-induced broadening and parton energy loss. Our numerical results show that the combined effect is positive $\gamma$-hadron pairs with small $p_{\rm T}^{\gamma}$ at RHIC energy in \autoref{fig:ATLAS_DphiRMS_ptGam}. For LHC energy and the large $p_{\rm T}$ $Z/\gamma$-hadron pairs at RHIC, for example $p_{\rm T}^{\gamma}>20$ GeV and $p_{\rm T}^{h}>2$ GeV, the combined effect becomes negative due to the stronger anti-broadening effect from parton energy loss than the medium-induced broadening effect with given the $\hat{q}$-median. However, the negative effect is so weak to be difficultly observed at LHC experiments.

\begin{figure}[!ht]
\centering
\includegraphics[width=1\linewidth]{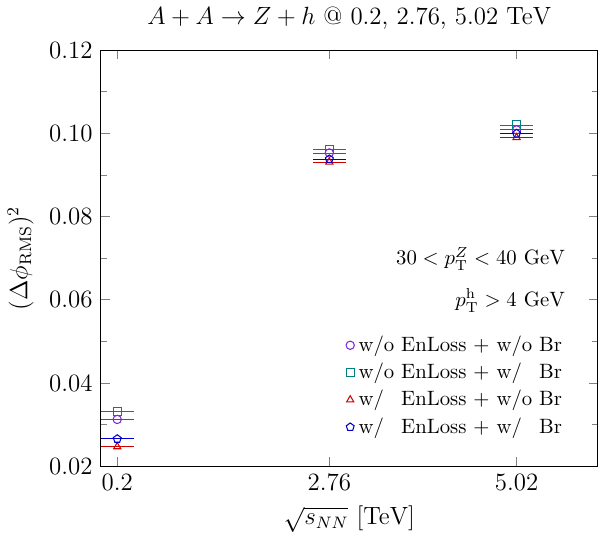}
\caption{(Color online) The RMS widths of angular correlation distributions of the $Z$-hadron pairs as a function of the colliding energy $\sqrt{s_{NN}}$ with or without the parton energy loss or the medium-induced broadening effects.}
\label{fig:ATLAS_DphiRMS_sNN}
\end{figure}

\begin{figure}[!ht]
\centering
\includegraphics[width=1\linewidth]{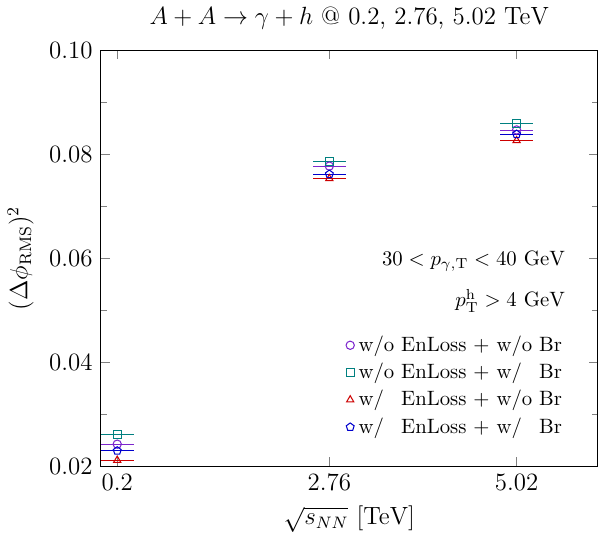}
\caption{(Color online) The RMS width of angular correlation distributions of $\gamma$-hadron pairs as a function of the colliding energy $\sqrt{s_{NN}}$ with or without the parton energy loss or the medium-induced broadening effects.}
\label{fig:ATLAS_DphiRMS_sNN_Gam}
\end{figure}

Now that we use the RMS width to quantify the dependence of the $Z/\gamma$-hadron angular correlation on the transverse momentum as well as the medium ``kick" and the energy loss, let's also check the dependence on the mass center energy in heavy-ion collisions. Shown in \autoref{fig:ATLAS_DphiRMS_sNN} for $Z$-hadron pairs and \autoref{fig:ATLAS_DphiRMS_sNN_Gam} for $\gamma$-hadron pairs are 
the RMS widths as a function of the mass center energy in central $AuAu$ collisions at $\sqrt{s_{\mathrm{NN}}}=0.2$ TeV and central $PbPb$ collisions at $\sqrt{s_{\mathrm{NN}}}=2.76$ TeV and 5.02 TeV, respectively. The corresponding $pp$ results are also shown for comparisons with the same cuts, $30<p_{\rm T}^{Z/\gamma}<40$ GeV and $p_{\rm T}^{h}>4$ GeV. We set the same three cases for $AA$ collisions as in 
\autoref{fig:ATLAS_DphiRMS_ptZ} and 
\autoref{fig:ATLAS_DphiRMS_ptGam}.
Numerical results for the $Z/\gamma$-hadron production show that the width of the angular distribution becomes larger with the increasing of the mass center energy in both $pp$ and central $AA$ collisions. The weakening angular correlation is resulted from the dominance of the Sudakov contribution which is stronger due to the higher scale $Q^2$, as shown in Eqs. (\ref{eq:sud_p}) and (\ref{eq:sud_np}). Comparing to the angular correlation in $pp$ collisions (purple circle), in $AA$ collisions the parton energy loss effect (red triangle) enhances the angular correlation while the medium-induced broadening effect (green square) makes the decorrelation.

Comparing the numerical results above, one can see that the width of the $\gamma$-hadron angular distribution is smaller than the corresponding width of the $Z$-hadron. This 
is because with the same kinematics cuts the 
$\gamma$-hadron process has a lower hard scale $Q^2$ than the $Z$-hadron process due to $Z$ mass, and thus encounters a weaker Sudakov effect.
When subtracting $\Delta \phi_{\rm RMS}^{pp}$ from $\Delta \phi_{\rm RMS}^{AA}$, the Sudakov effect almost cancels out, which makes $(\Delta \phi_{\rm RMS}^{AA})^2-(\Delta \phi_{\rm RMS}^{pp})^2$ similar between the $\gamma$-hadron and $Z$-hadron angular correlations.


\section{Summary}\label{sec:summary}

We have applied the Sudakov resummation improved pQCD parton model, including the Higher-Twist energy loss $\Delta E$ and medium-induced transverse momentum broadening $\left\langle p_{\perp}^2 \right\rangle$, to study  
medium modifications on momentum and angular correlations between a large transverse momentum hadron and the $Z/\gamma$ trigger in relativistic heavy-ion collisions.
The jet transport coefficient extracted by the IF-Bayesian inference is used to constrain $\Delta E$ as well as $\left\langle p_{\perp}^2 \right\rangle$.
Numerical results fit data very well for the suppression of the $Z/\gamma$-hadron yields as a function of $p_{\rm T}^{\rm h}$, $\xi_{\rm T}$, $z_{\rm T}$ or $\Delta \phi$ in $AA$ collisions at RHIC/LHC, which reveals a consistent description of jet quenching. 

The Sudakov resummation technique used in this study for the $Z/\gamma$-hadron production gets rid of the divergence of the back-to-back azimuthal angular distribution in both $pp$ and $AA$ collisions. 
This is, somehow, similar to that a $\delta$-function-like distribution is smeared to a Gaussian-like distribution. 
Different from $pp$ collisions, in $AA$ collisions the QGP medium induced broadening effect is approximately cumulated to the smearing effect \cite{Chen:2016vem,Chen:2016cof}. 
In another word, such a medium effect results in the decorrelation in the angular distribution for the $Z/\gamma$-hadron production in $AA$ collisions. 
However, jet energy loss which does not change the jet direction enhances the correlation in the angular distribution for the $Z/\gamma$-hadron production in $AA$ collisions. 
Due to the presence of jet energy loss in $AA$ collisions, with the same $p_{\rm T}$ cuts for the hadrons, the parent partons in $AA$ collisions have larger initial energy than that in $pp$ collisions.
In addition, the vacuum Sudakov effects increase with the partons’ initial energy. So jet energy loss not only leads to large $p_{\rm T}$ $Z/\gamma$-hadron suppression but also presents an anti-broadening in the angular distribution for the $Z/\gamma$-hadron production in $AA$ collisions.

We use the RMS width of the normalized angular distribution of the $Z/\gamma$-hadrons to analyze the competition between the positive broadening from the medium induction and the negative anti-broadening from jet energy loss.
Numerical results show that 
when the colliding energy or trigger transverse momentum is large, the vacuum Sudakov effect overpowers the medium-induced broadening effect, resulting in a weak positive broadening of the azimuthal distribution.
And when the associated hadron has a relatively low $p_{\rm T}^{\rm h}$, the energy loss mechanism results in a strong negative broadening.
Under this circumstance, the azimuthal angular distribution will display an anti-broadening phenomena, where the distribution for $AA$ seems to be sharper than that for $pp$ collisions.

Our numerical results demonstrate that low colliding energy with small trigger $p_{\rm T}$ is more suitable to study the angular decorrelation of two-particle correlations, and that both parton energy loss and medium-induced broadening should be considered simultaneously to provide an accurate description to the jet quenching effect in heavy-ion collisions. 


\begin{acknowledgement}
{\bf{Acknowledgements}}
We thank Guang-You Qin and Xin-Nian Wang for helpful discussions. This work is supported by Natural Science Foundation of China (NSFC) under Grant No. 11935007 and by the Guangdong Major Project of Basic and Applied Basic Research No. 2020B0301030008 and the Science and Technology Program of Guangzhou No. 2019050001. \\

\noindent
{\bf{Data Availability Statement}} This manuscript has no associated data or the data will not be deposited. [Authors’ comment: Data points used to draw the plots and predictions for the future measurements can be obtained through email upon request.]

\end{acknowledgement}




\printbibliography

\end{document}